\documentclass[reprint,amssymb, amsmath, aip,cha,twocolumn]{revtex4-1}
\usepackage{graphicx}
\usepackage[caption = false]{subfig}
\usepackage{color}
\usepackage{epsfig}
\usepackage{ifpdf}
\usepackage{url}%
\usepackage{multirow}
\usepackage{bm}
\usepackage[colorlinks=true,linkcolor=blue]{hyperref}%
\usepackage{cleveref}
\expandafter\ifx\csname package@font\endcsname\relax\else
 \expandafter\expandafter
 \expandafter\usepackage
 \expandafter\expandafter
 \expandafter{\csname package@font\endcsname}%
\fi
\begin{document}
\title{Excitation of dust acoustic shock waves in an inhomogeneous dusty plasma}%
\author{Garima Arora}%
\email{garimagarora@gmail.com}
\author{P. Bandyopadhyay}
\author{M. G. Hariprasad}
\author{A. Sen}
\affiliation{Institute For Plasma Research, HBNI, Bhat, Gandhinagar, Gujarat, India, 382428}%
\date{\today}
\begin{abstract}
 An experimental investigation of the propagation characteristics of shock waves in an inhomogeneous dusty plasma is carried out in the Dusty Plasma Experimental (DPEx) device. A homogeneous dusty plasma, made up of poly-dispersive kaolin particles, is initially formed in a DC glow discharge Argon plasma by maintaining a dynamic equilibrium of the pumping speed and the gas feeding rate. Later, an equilibrium density inhomogeneity in the dust fluid is created by introducing an imbalance in the original dynamic equilibrium. Non-linear wave structures are then excited in this inhomogeneous dusty plasma by a sudden compression in the dust fluid.  These structures are identified as shock waves and their amplitude and width profiles are measured spatially. The amplitude of a shock structure is seen to increase whereas the width broadens as it propagates down a decreasing dust density profile. A modified-KdV-Burger equation is derived and used to provide a theoretical explanation of the results including the power law scaling of the changes in the amplitude and width as a function of the background density.  
\end{abstract}
\maketitle
 \section{Introduction}\label{sec:intro}
 A dusty plasma, consisting of electrons, ions and micron/sub-micron sized charged dust particles, provides an excellent medium for the study of collective phenomena as it supports a wide variety of linear and nonlinear waves and coherent structures. During the past couple of decades, extensive theoretical and experimental studies have been carried out on the excitation of linear modes like Dust Acoustic Waves (DA) \cite{dustacousticwave}, Dust Ion Acoustic (DIA) waves \cite{pkshuklaDustionacousticWave}, Dust Lattice Waves \cite{latticewaves}, non-linear modes like Dust Solitary (DS) waves \cite{soilton,samsonovsoliton,senprecursor}, Dust Acoustic Shock (DAS) Waves \cite{heinrichshock, surbhishock, nakamurabow} and various coherent structures like voids\cite{voidGoree}, vortices \cite{vortexExcitationLaw} \textit{etc.}. Dust Acoustic Shock Waves (DASW) \cite{heinrichshock,nakamurabow,Usachevshock} constitute an important class  of non-linear waves that are frequently found to be excited in laboratory dusty plasmas as well as in astrophysical dusty plasmas \cite{steinolfson1993venus, gosling1967vela, vidotto2011shock}.  Shock waves are highly non-linear structures, which form with a characteristic sudden jump in any one of the physical parameters such as pressure, velocity, temperature and density. These nonlinear waves get excited when the dissipation (due to collisions or viscosity) in the medium plays a significant role along with non-linearity and dispersion. In dusty plasmas, dissipation can arise either from frequent dust-neutral collisions or dust-dust coupling effects. In a weakly coupled dusty plasma, the dissipation comes primarily from the kinematic viscosity due to frequent dust-neutral collisions and/or dust charge fluctuations, whereas for a strongly coupled dusty plasma the bulk and shear viscosity play a crucial role in providing dissipation. \par
Dust acoustic shock waves have been studied extensively by many researchers worldwide both theoretically\cite{theoretical_shock} and experimentally \cite{heinrichshock,nakamurabow,Usachevshock}. Samsonov \textit{et al}. reported shock formation in a Radio frequency (RF) produced 3D complex plasma under microgravity conditions in the PKE-Nefedov device\cite{samsonovshock}. Heinrich \textit{et al.} \cite{heinrichshock} observed self-excited dust acoustic shock waves in a direct current glow discharge dusty plasma that was generated when the dust cloud went through two slits. Nakamura \textit{et. al.} \cite{nakamurabow} and Jaiswal \textit{et al.} \cite{surbhishock} reported experimental observations of bow shock and oscillatory shock structures, respectively, in flowing dusty plasmas when the dust fluid was made to flow supersonically around a stationary charged object. It is to be noted that all these experiments were carried out in a homogeneous dusty plasma medium. However, in an experimental situation, the presence of density inhomogeneity is likely to be significant when one performs experiments in a large sized dusty plasma medium. In the literature there exist a very limited number of theoretical works devoted to the study of shock waves propagating in an inhomogeneous plasma medium \cite{xiao2005dust,zhang2019propagation}. They show that due to the presence of inhomogeneities in the plasma density, the relationships between amplitude, width and Mach number of shock waves deviate significantly from those obtained in a homogeneous medium.\par
Tadsen \textit{et al.} \cite{tadsen2017amplitude} performed experiments to investigate the dependence of the amplitude of spontaneously excited dust acoustic waves in an inhomogeneous plasma having spatial variations of dust charge, ion density, and dust density. To the best of our knowledge, no experiments have been done so far to examine the nature of propagation of dust acoustic shock waves in an inhomogeneous medium. In this paper we report on a first such experiment carried out by us to observe the temporal behavior of dust acoustic shock waves as they propagate in an inhomogeneous dusty plasma medium. 
Our experimental results show that the amplitude and width of these shock structures increase when they propagate down a decreasing density profile of the dusty plasma medium. The increase is inversely proportional to the decrease in the background density in a fractional power law manner. We provide a qualitative theoretical understanding of the experimental results by constructing a modified-KdV-Burger model equation that includes dissipation effects arising from both dust neutral collisions and dust-dust coupling induced viscosity. \\
The paper is organized as follows: in Sec.~\ref{sec:setup} the experimental set-up and details of the experimental  procedure are described. Sec.~\ref{sec:results} contains the experimental findings and a brief discussion on them.  A theoretical model to describe the experiments are discussed in Sec.~\ref{sec:model}. Sec.~\ref{sec: summary} provides a summary and some concluding remarks. \\
\section{Experimental Set-up and procedure}\label{sec:setup}
\begin{figure}[ht]
\includegraphics[scale=0.5]{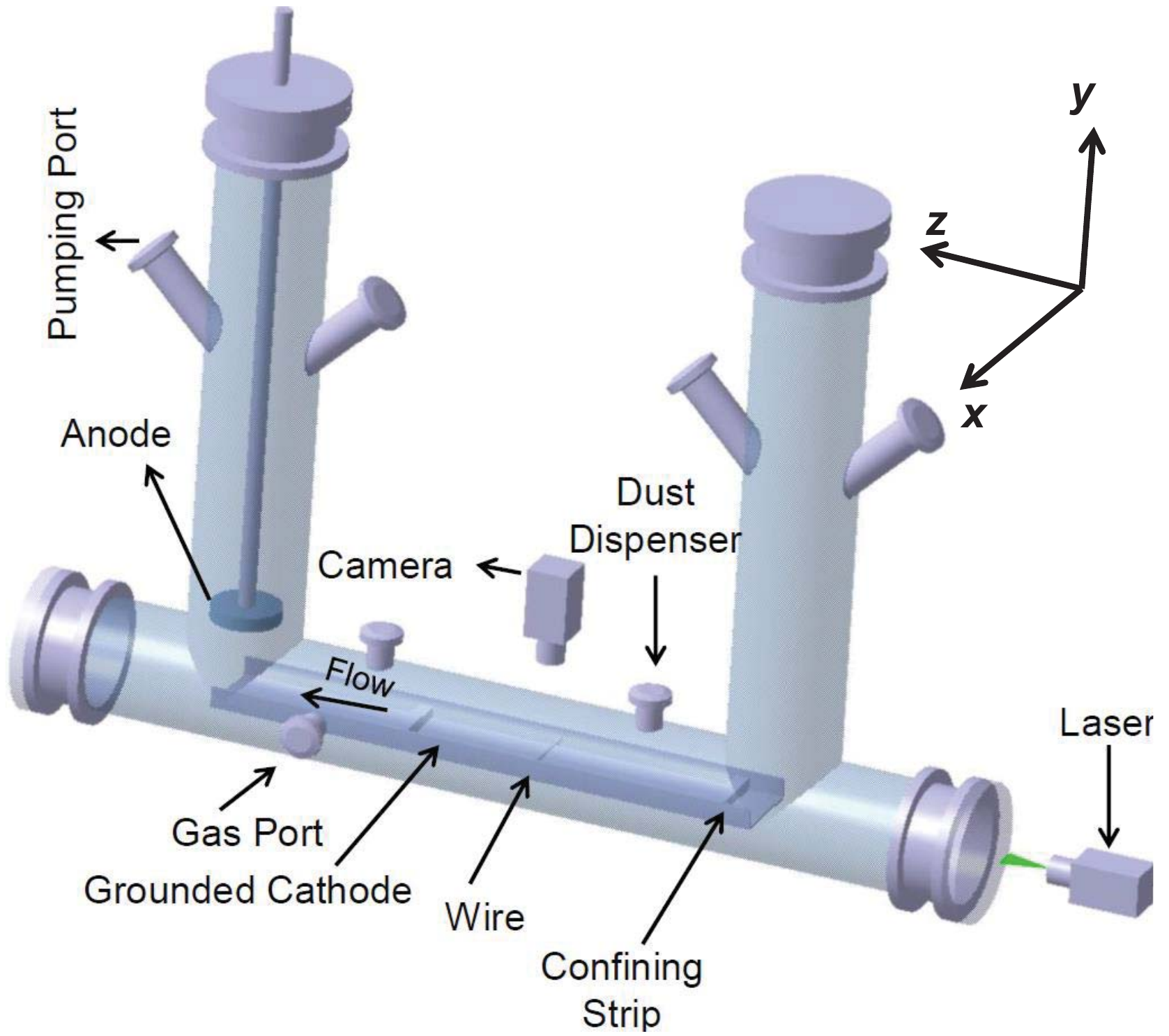}
\caption{\label{fig:fig1} A schematic diagram of dusty plasma experimental (DPEx) device.  }
\end{figure}
These experiments are performed in an inverted $\Pi$-shaped Dusty Plasma Experimental (DPEx) Device \cite{surbhirsi}. Fig.~\ref{fig:fig1} gives a schematic of the DPEx device showing the various radial and axial ports that are used for different experimental needs. The pumping port attached to a rotary pump through a gate valve is used to evacuate the chamber whereas the gas port is connected with a flow meter to feed Argon gas into the vessel in a controlled way. A disc shaped anode and a tray shaped grounded cathode serve as electrodes for the generation of a plasma whereas micron sized poly-dispersive particles of diameter 2-5 $\mu$m are used to create a dusty plasma. Two SS cuboid-shaped strips are placed on the cathode in order to confine the charged dust particles in the axial direction. One of these axial confinement strips is kept at the right edge of the cathode whereas the other one is placed at a distance of 28 cm from the first one. The upturned sides of the 6~cm wide cathode tray provide radial confinement. A biased copper wire is mounted radially on the cathode (approximately 20 cm away from the right edge of the cathode) to confine the dust particles axially between the right confining strip and the wire. For a more detailed description of the device and the attached diagnostics the reader is directed to Ref.\cite{surbhirsi}. In the present set of experiments, the copper wire is always kept at the ground potential. A  Direct Current (DC) power supply is used for the production of a glow discharge Argon plasma. \par
To start with, the chamber is evacuated with the help of the rotary pump to attain a base pressure of p $\sim$ 0.1 Pa and later the working pressure is set to  p $\sim$ 11 Pa by adjusting the dynamic equilibrium of the pumping speed and the gas flow rate. In this configuration, the gate valve is opened at $\sim 20\%$ whereas the gas flow meter is opened at 5$\%$. An Argon discharge plasma is then initiated by applying a DC voltage $V\sim$ 330~V at this working pressure. Plasma parameters such as plasma density ($n_i$) $\sim 1.5\times10^{15}$ $/m^3$, electron temperature ($T_e) \sim$ 4 eV, floating potential ($V_f$) $\sim 290$~V and plasma potential ($V_p$) $\sim 310$~V with respect to the grounded wire are measured, in the absence of dust particles, by using a single Langmuir probe and emissive probe. The measurement techniques and the axial profiles of different plasma parameters over the range of discharge conditions are presented in detail in Ref.~\cite{surbhirsi}.  As soon as the plasma is initiated, the dust particles of radius ($r_d$) $\sim 1-2.5$~$\mu$m that are sprinkled on the cathode tray prior to the discharge, get negatively charged and levitate \textcolor{black}{in the cathode sheath region} by a balance of the sheath electrostatic force and gravitational force in the vertical direction. {\color{black} The vertical height of the cloud bottom varies in the range of 1.3 cm to 1.7 cm depending on the discharge conditions, whereas the vertical width of the cloud is always between $1~cm$ to $1.2$~cm.} The repulsive interactions among the negatively charged dust particles in the radial direction can be nullified by the strong radial confinement force provided by the cathode edges.  However, the repulsive interaction in the axial direction is taken care of by the grounded copper wire and the right strip. The force balance in axial, radial and vertical directions creates an equilibrium dust cloud. The average mass of the levitated dust particles is $m_d\sim 8.8\times 10^{-14}$ kg and the charge ($Q_d$) acquired by these micro particles at this discharge condition is $\sim 10^4$e estimated from a Collision Enhanced Plasma Collection (CEC) model \cite{khrapakcec2,khrapakcec} for particles of average radius $2.0~\mu$m. Depending upon the mass ($m_d$) and charge ($Q_d$) of the particles and the vertical cathode sheath electric field ($E(y)$), the micron-sized polydispersive particles levitate at different heights by balancing the gravitational force (acting in the downward direction) and the electrostatic force (acting in the upward direction for the negatively charged dusty particles). Since $m_d \propto r_d^3$ and $Q_d \propto r_d$ (provided $T_e$ remains constant), one can conclude from the force balance condition that the particles of the same size should levitate at the same height, $y$, provided the electric field remains the same over the axial extent at that height. {\color{black} We have also experimentally confirmed that $E(y)$ (which corresponds to the slope of the plasma potential) and  $T_e$ remain nearly constant at different axial locations for a given set of discharge conditions. This is in agreement with our earlier observations as reported in \cite{surbhirsi}.} A perfectly aligned parallel thin laser sheet (in x-z plane) \textcolor{black}{of width 1~mm  illuminates a central slice of the dust cloud consisting of particles of the same size for a given set of discharge conditions.} The dynamics of the dust particles are captured by a fast CCD camera having a frame rate of 200~frames/s with a spatial resolution of 42 $\mu$m/pixel. \par
 \begin{figure}[ht]
\includegraphics[scale=0.65]{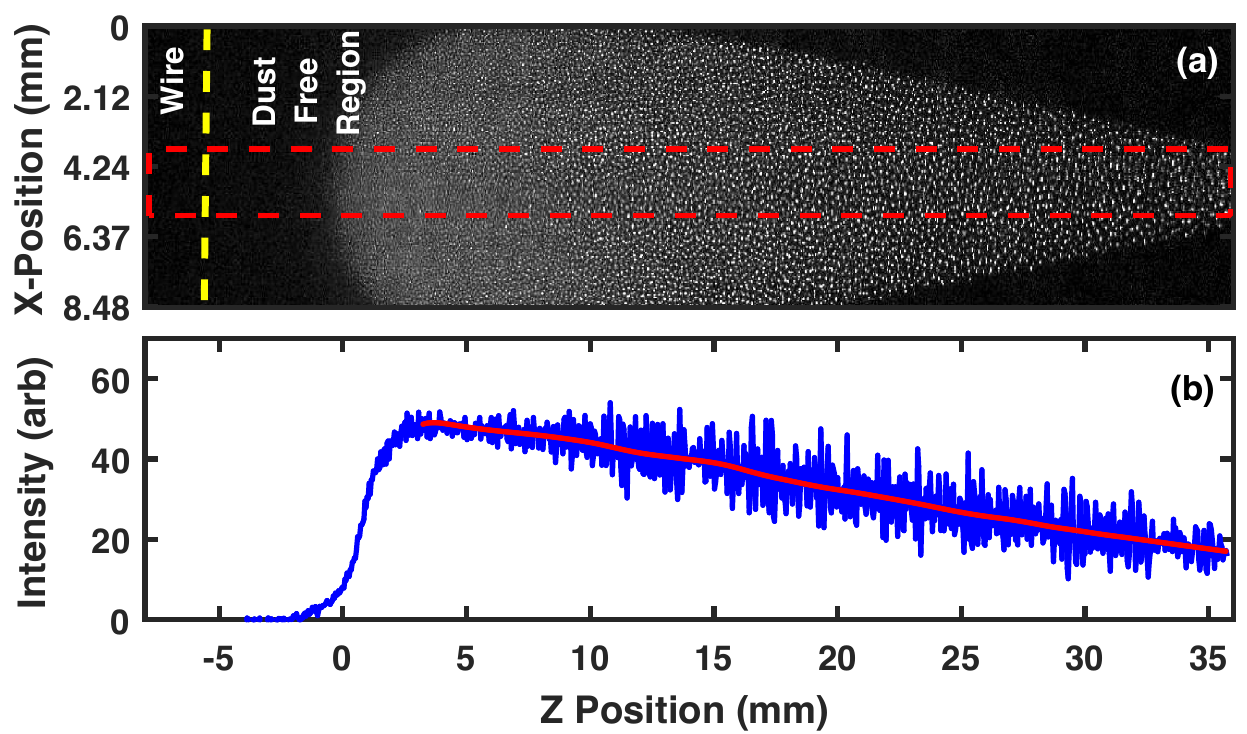}
\caption{\label{fig:fig2} (a) An experimental image of equilibrium dust cloud with linear decreasing dust density. (b) Intensity profile extracted from fig. \ref{fig:fig2} (a).}
\end{figure}         
 Since, these experiments are aimed at investigating the propagation characteristics of non-linear waves in an inhomogeneous dusty plasma, {\color{black} we adopt the following sequence of procedures. First a uniform dusty plasma is created between the sheaths of the grounded wire and the right confinement strip by maintaining a dynamic equilibrium of the pumping speed and the gas feeding rate. We then alter this equilibrium by changing the pumping speed to create a pressure gradient. This causes a neutral flow towards the grounded wire and a concomitant flow of the dust particles due to the neutral drag force \cite{surbhiflow,garimadynamics} leading to an accumulation of the dust particles near the wire. Thus a second dynamic equilibrium state is reached that is maintained by the pressure gradient and the electrostatic repulsion of the sheath around the wire. The whole cloud is still confined between the wire and the right strip due to the sheaths at the two ends but now has a nonuniform distribution of particles. Fig.~\ref{fig:fig2}(a) displays such a dynamic equilibrium state of the dust cloud with a clear indication of the dust density gradient,} where the number of particles are more near the grounded wire and less towards the edge of the cathode. It is to be noted that the dust density is inferred from the pixel intensities of an image even in the case of polydisperse particles, provided the size of the dust particles in a particular layer remains the same. This is a standard technique used in the past by several researchers \cite{merlino2012nonlinear,heinrichshock}. As mentioned in Sec.~\ref{sec:setup}, the particles of same size from the range of polydispersive particles levitate in the particular layer by the balance of gravitation force and the cathode sheath electric force. As we are shining the laser in a particular x-z plane by keeping y-position fixed, it essentially means that we are capturing the scattered light mostly from monodisperse particles that exist in a particular layer. In Fig.~\ref{fig:fig2}(a) the dust cloud near the wire is not constituted of smaller dust particles; instead, the number of particles (of the same size) per unit volume is maximum there. Due to the higher dust density in that region, the camera is unable to identify them as separate particles. \par
The intensity profile of equilibrium dust cloud and the corresponding fitted curve is shown in fig.~\ref{fig:fig2}(b). The pixel intensity in fig.~\ref{fig:fig2}(b) is plotted by averaging the intensities of 50 pixels in the vertical direction. The area for which the average intensity is calculated is shown by a dashed rectangle in fig. \ref{fig:fig2}(a). This rectangle essentially shows that all the dark spaces (outside the rectangle) of the image does not contribute to the calculation of average pixel intensity. The region of interest is fitted by the solid line in fig.~\ref{fig:fig2}(b), where the intensity decreases monotonically if one goes to the right away from the wire. It is worth mentioning that the laser light is kept perfectly parallel to the cathode so that the light scattered from dust particles is not re-scattered from the other planes of the dust cloud. We have taken care that the camera does not saturate while capturing the images of the equilibrium dust cloud and the excitations of dust acoustic shock waves. We have also verified that the image intensity is linear with the dust number density by checking that the camera has a linear response with no offset.\par
\begin{figure}[ht]
\includegraphics[scale=0.95]{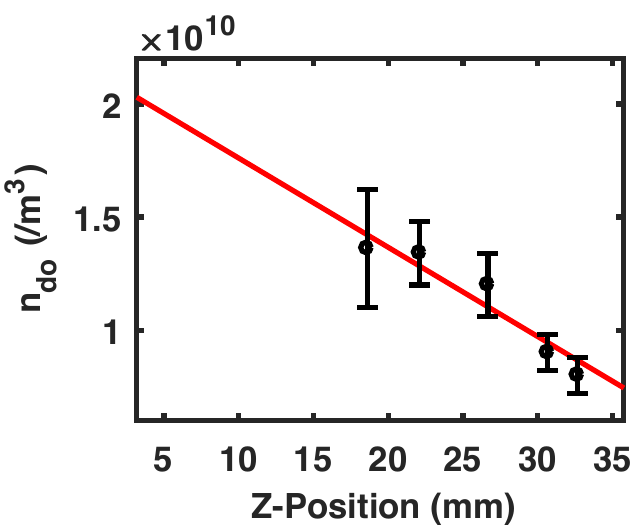}
\caption{\label{fig:fig7} Equilibrium dust density (solid line) constructed from the intensity profile as shown in Fig.~\ref{fig:fig2} and few measurements of dust density (closed circles) at $z \sim 18-32$~mm.} 
\end{figure}  
The equilibrium spatial density profile is constructed with the help of the intensity profile  shown in fig.~\ref{fig:fig2}(b) and a few measurements of dust density in the tail part of the dust cloud ($z \sim 18-32$~mm). The particles are well resolved in this spatial region, which allows us to measure the dust density (see the closed circles in fig.~\ref{fig:fig7}). {\color{black} Dust density ($n_{do}=3/4\pi d^3$) is estimated from the information of inter-particle distance ($d$) which is defined as the average distance from the reference particle to the neighboring particles. We have selected a reference particle at a particular axial location and measured the distance to all its neighboring particles and calculated the average value. To improve the statistics in the measurements, again a new reference particle is chosen at the same axial location but a different $x$ position, and the same procedure is performed to get the statistical mean and standard deviation. } {\color{black} The statistical error on the measurements of the dust density at the axial location 18-26 mm  are quite high since the resolution is very poor.}  By assuming the linear response of the camera and using the measured values of dust density near $z \sim 18-32$~mm, the intensity profile (solid line of fig.~\ref{fig:fig2}(b)) is then calibrated to dust density profile (see the solid line in fig.~\ref{fig:fig7}) in the region where the light intensity decreases linearly. As shown in fig.~\ref{fig:fig7}, the equilibrium dust density falls nearly monotonically from $\sim 2\times10^{10}$/m$^3$ to $\sim 0.7\times10^{10}$/m$^3$. \par
As a third step, shock waves are excited in this nonuniform dusty plasma medium by creating a sudden jump in the dust density. The compression in the dust density near the grounded wire is generated by momentarily flowing the dust fluid from right to left using a Single Gas Injection (SGI) technique. In this technique, {\color{black} the dynamic equilibrium is further perturbed by a short pressure pulse to excite the dust density waves. The pressure pulse is generated by tweaking the gas flow at the Gas Port (as shown in Fig.~\ref{fig:fig1}) for a short interval ($\sim 500~ms$).} Due to the abrupt pressure drop near the gas port, the neutrals rush axially towards the pump  and carry the dust particles along from right to left \cite{garimadynamics}. When the gas flow rate is restored to its initial value the dust particles come back to their original positions. The details of this technique of flow generation in the dust fluid and measurements of the flow velocity are available in greater detail in Ref.\cite{surbhiflow}. 
As discussed above, when the gas flow rate is restored to its initial value, the dust particles come back to their initial positions and the dust cloud takes the original shape. The short intense density perturbation gives rise to a large amplitude propagating dust acoustic wave which takes the form of a nonlinear propagating wave train of shock structures - an asymmetric waveform of connected saw-teeth structures \cite{wmtham1974linear}.  The wave forms propagate an average distance of 30-40 mm over  0.20-0.35~s with a velocity ranging from 5-12 cm/s depending upon their amplitude and width. For the present set of discharge conditions, the dust neutral collision frequency ($\nu_{dn}$) comes out to be $\sim 11$~Hz \cite{epstein}, whereas the wave frequency of linear dust acoustic waves is measured to be $\sim 21$~Hz. However, it is to be noted that the wave is sustained for approximately 1.5 to 2~s after the initial excitation by the gas which is almost one order higher than the damping time ($2/\nu_{dn}$) $\sim 180$~ms, for a linear wave. Partly this is due to the nonlinear nature of these waves and the growth they experience as they travel down the density gradient but could also be possibly due to the additional energy drive from streaming ions present in the experiment. The details of these excited nonlinear structures and their propagation characteristics in the in-homogeneous dust density are discussed in Sec.~\ref{sec:results}. \par
\section{Experimental Results}\label{sec:results}
\begin{figure}
\includegraphics[scale=0.75]{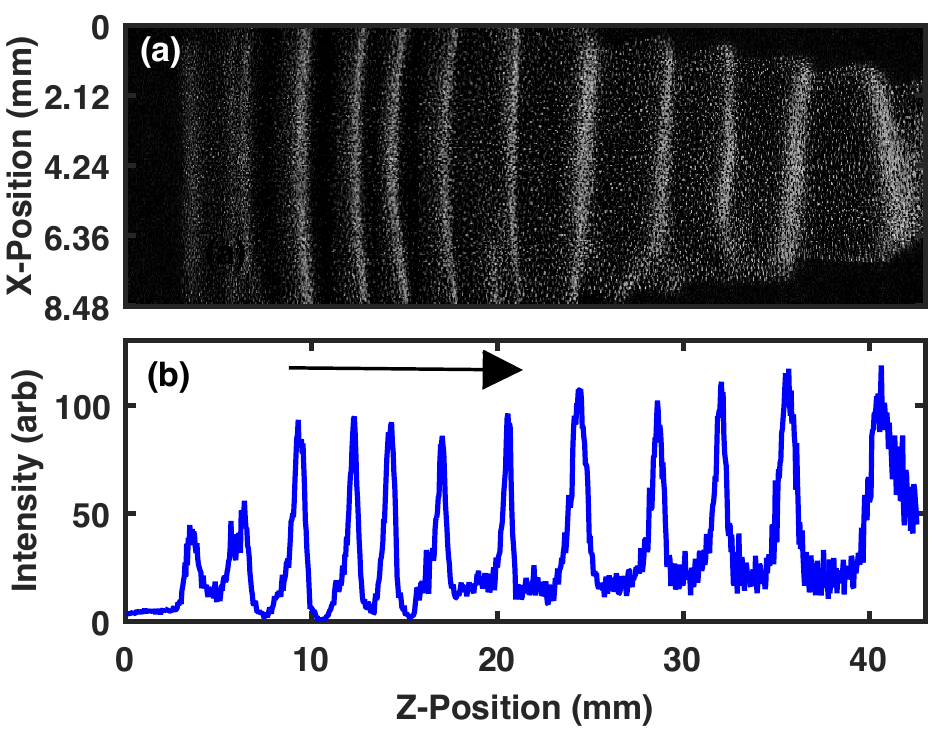}
\caption{\label{fig:fig3} (a) Experimental image of density crests in an inhomogeneous dust cloud. (b) Intensity profile of high amplitude density crests extracted from fig. \ref{fig:fig3}(a).  The arrow represents the direction of propagation of the shock fronts.}
\end{figure}
A typical snapshot image of the series of excited Dust Acoustic Shock Waves (DASWs) in an inhomogeneous dusty plasma is shown in Fig.~\ref{fig:fig3}(a), whereas Fig.~\ref{fig:fig3}(b) represents the corresponding intensity profiles of the density crests of DASWs. One can see clearly from both the subplots that these structures have different amplitudes and widths at different axial locations (along the Z-direction). In addition, it can also be seen that the distance between the crests of the structures keeps on increasing as one moves away from the wire. \textcolor{black}{The sharpness of the peaks, the high compression factor, the saw-teeth nature of their individual density profiles and their high propagation speeds ($\sim 5-12$ cm/s) compared to the dust acoustic speed ($\sim 1.5-3.5$ cm/sec) indicate that these structures are indeed non-linear shock formations.} To examine the symmetricity of the wavefronts, we have calculated a symmetric parameter, defined as the ratio ($R$) of the half widths at half maximum. {\color{black}We have chosen five wavefronts widely separated and located at a different axial location for a given frame and then estimate R  for these wavefronts for 20 frames.} For our experimental wave forms, the value of $R$ comes out to be $\sim 1.204-1.402$, whereas its value varies within $\sim 0.912-1.024$ for the case of linear dust acoustic waves \cite{choudhary2016propagation} and $\sim 0.954-1.006$ for the case of non-linear dust acoustic waves \cite{merlino2012nonlinear, williams2012spatial}. \textcolor{black}{In contrast to these linear and nonlinear waves, the value of $R$ for dust acoustic shock waves yields a higher value of $R\sim 1.62-2.05$ over time \cite{heinrichshock}. It essentially ensures that the structures excited in our experiments are indeed asymmetric in shape and significantly deviate from the conventional symmetrical structures observed in the excitation of linear or nonlinear dust acoustic waves like solitons and cnoidal waves.} These nonlinear dust acoustic saw-teeth structures are formed due to a sudden change in the dust density near the wire.
\begin{figure}
\includegraphics[scale=0.7]{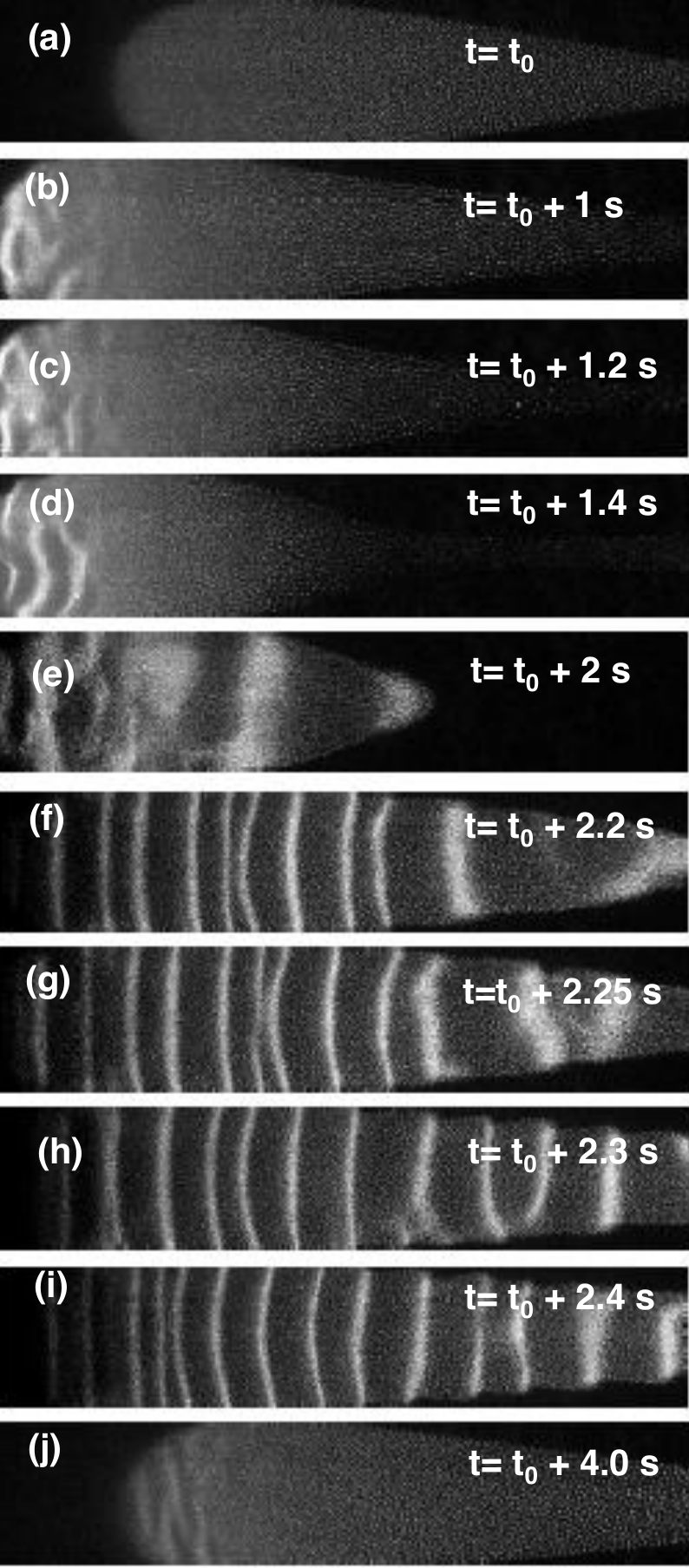}
\caption{\label{fig:fig_new} Sequence of images: (a) Stable unperturbed inhomogeneous dust cloud, (b--e) spontaneously excited dust acoustic waves in the compressed dust cloud, (f--i) excitation and propagation of nonlinear shock waves and (j) image of dust cloud when the gas flow rate is restored. }
\end{figure}
They actually constitute a nonlinear traveling wave with steepened crests - in other words a wave train with saw-teeth shaped periodic (shock \cite{wmtham1974linear}) structures. This is the nonlinear state of a dust acoustic plane wave excited by the perturbation induced by the act of density compression near the wire and its subsequent relaxation. The sequence of events is as follows (see Fig.~\ref{fig:fig_new}): In the initial stage (depicted in Fig.~\ref{fig:fig_new}(a)) when the inhomogeneous dust cloud is not perturbed there is no spontaneous excitation of any dust acoustic waves (DAWs). Fig.~\ref{fig:fig_new}(b)--(e) show that the nonlinear DAW excitation occurs after the dust density is momentarily compressed near the wire to create a large density perturbation and as this perturbation moves down the density gradient. Once excited the nonlinear DAW continues to be generated and to travel from left to right at a speed a few times larger than the dust acoustic speed as shown in Fig.~\ref{fig:fig_new}(f)--(i). {\color{black} Fig.~\ref{fig:fig_new}(j) shows that the dust cloud takes approximately its original shape and position (other than the excitations) when the gas flow rate is restored.} We do not fully understand what causes the continuous  generation of the wave after the initial perturbation ceases but we do observe them experimentally for a long time. A possible continuous driver of these waves could be the ion streaming which is always present in our experiment \cite{hari2018} because of asymmetric configuration of the electrodes. The streaming of ions can be the energy source sustaining the nonlinear wave propagation once the pressure perturbation has initiated the large amplitude DAW. It is worth mentioning that these nonlinear structures are not solitary waves as they do not maintain a constant solitonic parameter ($Amplitude \times Width^2$) over time as seen in earlier experiments of Samsonov \textit{et al.} \cite{samsonovsoliton}. Instead, one can see from Fig.~\ref{fig:fig3}(a) and (b) that the higher amplitude density crests get excited with higher width, which clearly show that these excited structures are not Dust Acoustics Solitary Waves. \par
\begin{figure}[ht]
\includegraphics[scale=0.8]{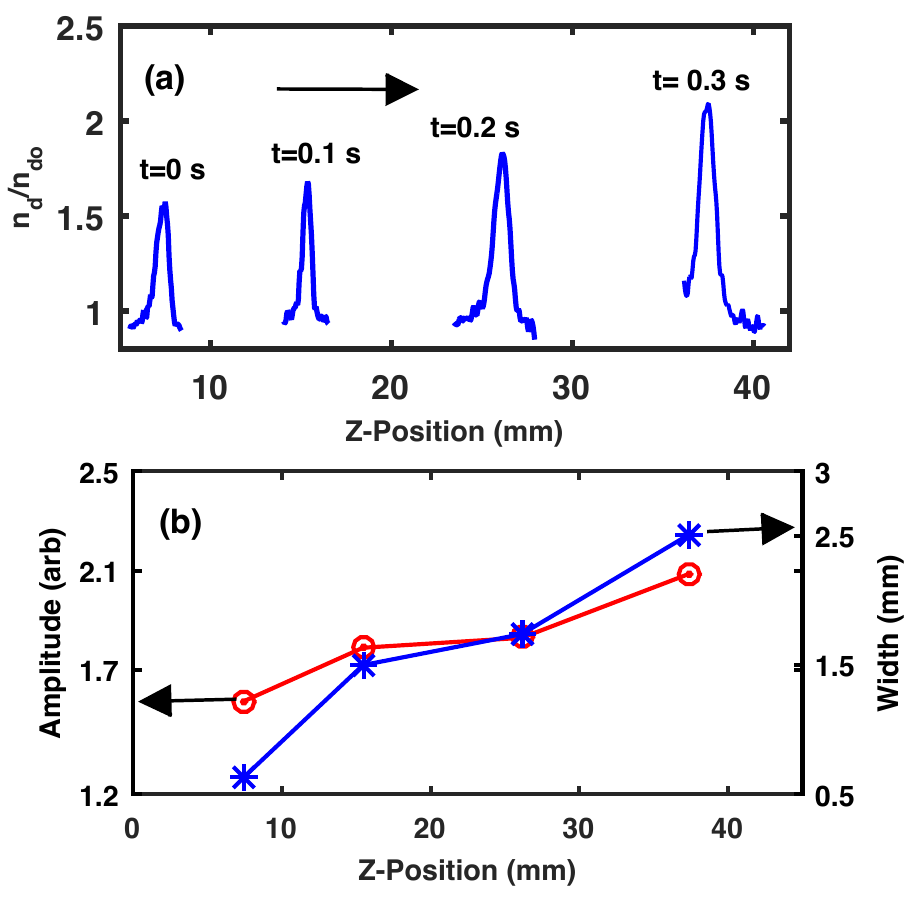}
\caption{\label{fig:fig5} (a) Time evolution of shock wave front in space. (b) Variation of amplitude and width of that particular shock front in space as shown in Fig.~\ref{fig:fig5}(a). }
\end{figure}
To study the modification in the propagation characteristics  due to the inhomogeneity in the equilibrium dust density, we carry out a detailed analysis of the evolution of the dust acoustic shock waves when they are excited in an inhomogeneous dusty plasma.  Figure \ref{fig:fig5}(a) shows the time evolution of one of the shock fronts in intervals of 0.1~s. The shock front is represented by the density compression factor,
defined by the standard expression $n_{d}(z)/n_{do}(z) \sim I(z)/I_o(z)$\cite{annibaldi2007dust, heinrichshock}, where $n_d(z)$ is the instantaneous dust density and $n_{do}(z)$ is the background equilibrium dust density at the same $z$-location, whereas $I(z)$ and $I_o(z)$ are the intensities of the perturbed dust cloud and  background equilibrium dust cloud, respectively. It is clear from the figure that the shock front propagates from left to right (i.e. away from the wire) and while propagating, the compression factor $n_d/n_{do}$ of that particular shock front increases due to the decrease in the dust density. Figure \ref{fig:fig5}(b) shows a quantitative analysis of the amplitude and width of shock wave front while it propagates. One can clearly see that the amplitude of a particular shock front increases and width broadens up in the course of propagation. In addition, as seen in Fig.~\ref{fig:fig5}(a), the wave front progressively travels longer distances (7.90~mm, 10.73~mm and 11.4~mm) for a given time duration of 0.1~s. It essentially indicates that the wave front propagates with a higher velocity as it moves through a lower density medium. {\color{black} As is well known from theoretical and experimental studies of nonlinear waves like cnoidal waves, solitons etc., the speed of a nonlinear wave depends on its amplitude \cite{pintudasw,bailung,rao1994nonlinear,Usachevshock}. The change in amplitude is inversely proportional to the equilibrium values of both the dust charge and the dust density  as discussed in past theoretical studies \cite{singh1998linear,singh1999effect,zhang2019propagation}. Thus the existence of a density gradient and charge gradient can change the amplitude and thereby change the velocity of the wave. The two gradients are related - a dust density gradient can lead to a dust charge gradient \cite{khrapakcec}. In our experiments, the dust density drops from 2$\times10^{10}$ to 0.7$\times10^{10}$~$/m^3$ over a distance of 35 mm. Using the Collision Enhanced Plasma Collection (CEC) model \cite{khrapakcec} we have calculated the corresponding change in dust charge over the same distance and found it to change from 1.5$\times10^{-15}$ to 1.53$\times10^{-15}$~C. Thus while the percentage change in dust density is $\sim 65\%$  the corresponding change in charge is only $\sim 2\%$. Therefore the influence of the dust charge density can be taken to be negligible and the principle effect on the amplitude comes from the density gradient  - it increases as it travels towards a region of lower density and consequently its speed increases. This is consistent with the experimental observation.}\par
\begin{figure}[ht]
\includegraphics[scale=0.8]{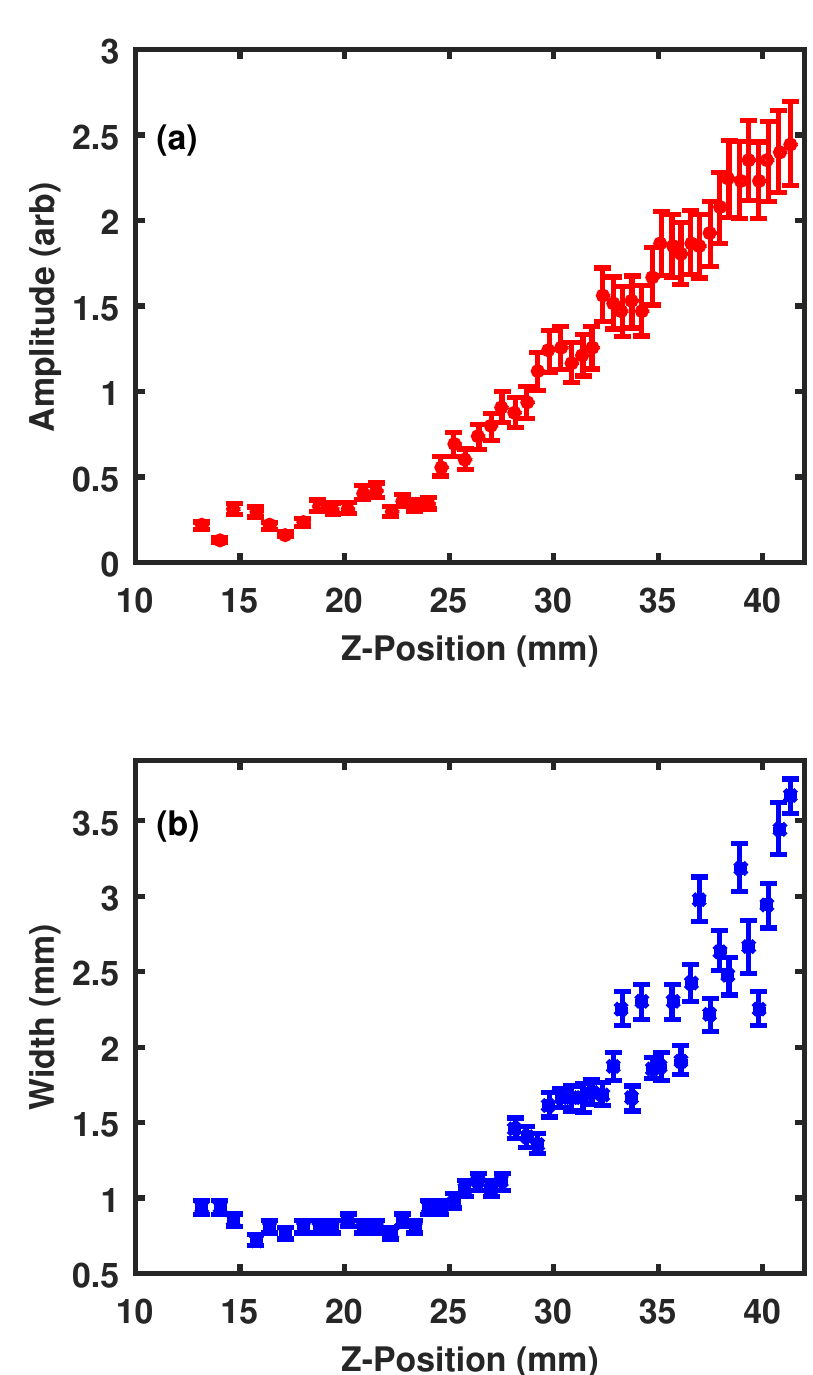}
\caption{\label{fig:fig6} Variation of (a) amplitude and (b) width of shock fronts along z-position. }
\end{figure}  
The shock parameters, namely the amplitude and the width are obtained by following the technique used by Heinrich \textit{et al.} \cite{heinrichshock} and Annibaldi \textit{et al.}\cite{annibaldi2007dust}. For a given $z$, the amplitude is estimated by $n_d/n_{do}-1$, whereas the shock width (or thickness) is defined as the difference between the steep edge point and the peak point of a shock front. It is to be noted that all the shock fronts are considered for different experimental shots and their amplitudes and widths are calculated by binning the space along the axial direction and this is plotted in Fig.~\ref{fig:fig6}. It shows that as the shock waves propagate from higher density (nearby the wire) to lower density (far away from the wire), both the amplitudes (see Fig.~\ref{fig:fig6}(a)) and the widths (see Fig.~\ref{fig:fig6}(b)) follow the same increasing trend. These present findings are in contrast to a homogeneous dusty plasma where it is found that both the amplitude and width of shock waves decrease in course of time due to the presence of strong dissipation in the medium \cite{heinrichshock}. However in our experiments, the wave train with its large amplitude (due to the manner of its excitation) is highly nonlinear and experiences both growth (due to propagation in a decreasing density gradient) and damping due to dust-dust correlations (viscosity) and dust-neutral interactions (collisions). The combination of these factors lead to a propagating steady state wave form with the characteristics of an asymmetric (steepened) wave form that we observe in the experiments. The increasing profile of shock amplitude is in agreement with past theoretical predictions of Zhang {\it et al.} \cite{zhang2019propagation}.  However, the increasing trend of shock thickness (width) with the density inhomogeniety has not been observed or studied earlier.\par
\begin{figure}[ht]
\includegraphics[scale=0.8]{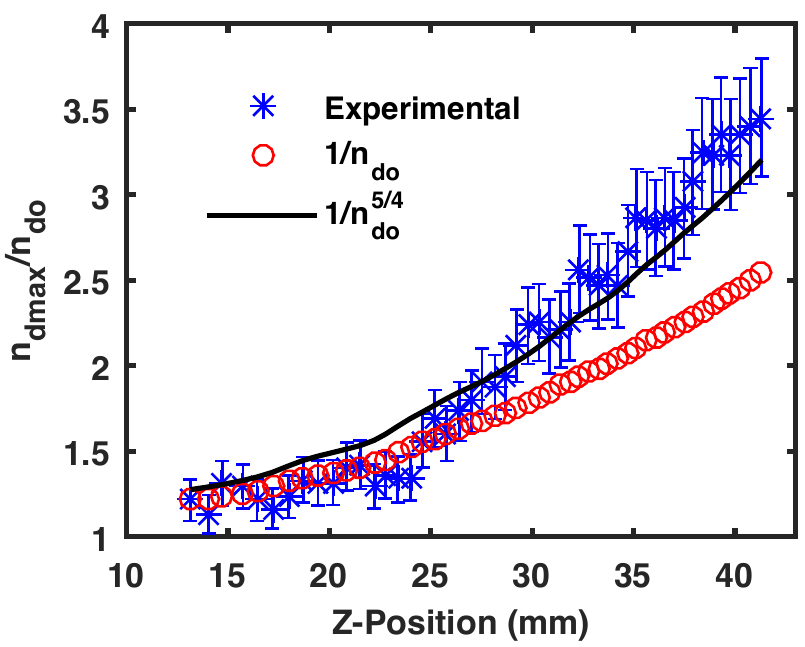}
\caption{\label{fig:fig8} Spatial variation of compression factors obtained experimentally and theoretically.  }
\end{figure}
In order to distinguish between the inhomogeneity effects on a linear perturbation and that on a nonlinear structure like our present shock structures we have compared the plot of the experimental compression factor with that of a hypothetical linear compression factor as it propagates in an inhomogeneous medium. The compression factor of a linear perturbation would increase as $n_d^{-1}$ by the very definition of the factor. This is shown in Fig~\ref{fig:fig8} where the  curve with \lq o' symbol represents the estimated compression factor of a linear wave propagating down a density gradient. The \lq*' symbol shows the experimentally obtained compression factor of the shock fronts at different spatial locations and the trend shows a significant deviation from the linear result thereby establishing that the propagation characteristics of the dust acoustic shock wave is distinctly different in an inhomogeneous medium as compared to a linear wave.  
The solid line fitted to the experimental data shows the scaling of the peak amplitude ($n_{dmax}/n_{do}$) to be
proportional to $1/(n_{do})^{5/4}$. The amplitude scaling is similar to that predicted theoretically by Singh \textit{et al.} \cite{singh1998linear, singh1999effect} for nonlinear dust acoustic waves. 
\section{Theoretical Modelling}\label{sec:model}   
To model our experimental results, we have derived analytically a modified KdV-Burger equation in an inhomogeneous dusty plasma. As discussed in Sec.~\ref{sec:setup} and Sec.~\ref{sec:results}, our experimental conditions ensure that there are no significant axial variation of dust size and dust charge in the dust layer under investigation. We have therefore only considered the effect of dust density gradient in the theoretical model to study the propagation characteristics of dust acoustic shock waves in an inhomogeneous dusty plasma.
Furthermore the experimental results pertain to shock front propagation in the axial direction only and so our theoretical model is restricted to one dimensional wave motion. For the dust acoustic wave dynamics the inertia of electrons and ions can be neglected as they are lighter than the dust component. Under these conditions, the electrons and ions are governed by Boltzmann distributions defined by their respective temperatures $T_e$ and $T_i$ as:
\begin{eqnarray}
n_e=n_{eo}exp(\frac{e\phi}{T_e}),\\
n_i=n_{io}exp(\frac{-e\phi}{T_i}),
\end{eqnarray}
where $\phi$ is the electrostatic potential. It has been reported in the earlier experiments of Jaiswal \textit{et al.}\cite{surbhirsi} that the plasma density remains nearly constant along the axial direction. Hence, the equilibrium electron ($n_{eo}$) and ion ($n_{io}$) densities are taken to be constant in the theoretical model. In the theoretical model, the dust particle motion is assumed to be adiabatic and as a result, we use the $\gamma$-model to express the dust pressure ($p_d$) as $\nabla p_d=\gamma_dT_d\nabla n_d$, where $\gamma_d, ~T_d$ and $n_d$ are the adiabatic constant, the dust temperature and the instantaneous dust density respectively The governing fluid equations for such a system can be written as:
 \begin{eqnarray}
\frac{\partial n_d}{\partial t}&+&\frac{\partial(n_dv_d)}{\partial x}=0,\label{eq:con}\\
\frac{\partial v_d}{\partial t}&+&v_d\frac{\partial v_d}{\partial x}-\frac{Z_de}{m_d}\frac{\partial \phi}{\partial x}+\nu_{dn}v_d+\gamma_d v_{td}^2\frac{1}{n_d}\frac{\partial n_d}{\partial x}=\frac{\eta_l}{n_d}\frac{\partial^2 v_d}{\partial x^2},\label{eq:mom}\\
\frac{\partial^2 \phi}{\partial x^2}&+&4\pi e(n_i-n_e-Z_dn_d)=0,\label{eq:pois}
\end{eqnarray}
where $v_{td}=\sqrt{\frac{T_d}{m_d}}$ and $n_d,v_d,\phi, m_d, Z_d,v_{td}$ are the density,  velocity, electrostatic potential, mass, charge, and thermal velocity of the charged particles respectively. {\color{black} The dust velocity $v_d$ used in the continuity and momentum equations (in Eqns.~\ref{eq:con} and \ref{eq:mom}) is the instantaneous velocity of the perturbed dust fluid element - the perturbation that leads to the formation of the nonlinear dust acoustic wave. In the model calculation we have ignored the gas flow velocity assuming it to be much smaller than the dust acoustic speed \cite{singh1998linear,singh1999effect}}. The other parameters come from the background properties of the medium with $\nu_{dn},~\eta_l,~n_i,~n_e,$ representing the dust-neutral collision frequency, viscosity and the background electron and ion densities respectively. The coupling between the dust particles provides bulk viscosity \cite{kaw1998low} in the medium which plays the role of dissipation. Due to this reason, the strong coupling induced viscosity is included in the momentum equation. In addition, the effect of frequent dust neutral collisions is also considered in the momentum equation (Eq.~\ref{eq:mom}) through the inclusion of the fourth term, by which the waves experience collisional damping while propagating.\par
A suitable set of stretched coordinates for the inhomogeneous plasmas can be defined by
\vspace*{-0.05in}
\begin{equation}
\xi=\epsilon^{1/2}\left(\int_{}^{x}\frac{dx^\prime}{\lambda(x^\prime)}-t\right),\hspace*{0.2in}\eta=\epsilon^{3/2}x,
\end{equation}
where $\epsilon$ is the smallness parameter and $\lambda$ is the velocity of the moving frame which can be determined self consistently $\eta_o=\epsilon^{1/2}\eta_l$ and $\nu_{dn}=\epsilon^{3/2}\nu_o$. Since we are considering only spatial gradients so,
\begin{eqnarray}
\frac{\partial n_{do}}{\partial \xi}=0;
\hspace*{0.1in}\frac{\partial \lambda}{\partial \xi}=0;
\hspace*{0.1in}\frac{\partial \phi_o}{\partial \xi}=0;
\hspace*{0.1in}\frac{\partial v_{do}}{\partial \xi}=0.
\end{eqnarray}
Using equations (6)--(7) into equations (3)--(5) we obtain the continuity equation in the form,
\begin{equation}
-\lambda\frac{\partial n_d}{\partial \xi}+\frac{\partial (n_dv_d)}{\partial \xi}+\epsilon \lambda \frac{\partial (n_dv_d)}{\partial \xi}=0,
\end{equation}
and the momentum equation becomes, 
\begin{widetext}
\begin{equation}
\begin{split}
-\lambda^2\frac{\partial v_d}{\partial \xi}+\lambda v_d\frac{\partial v_d}{\partial \xi}+v_d\epsilon\lambda^2\frac{\partial v_d}{\partial \eta}
-\frac{Z_de\lambda}{m_d}\frac{\partial\phi}{\partial \xi}-\frac{Z_de\lambda^2\epsilon}{m_d}\frac{\partial \phi}{\partial \eta}+ 
\frac{\lambda \gamma_d v_{td}^2}{n_d}\frac{\partial n_d}{\partial \xi}+\frac{\lambda^2\epsilon\gamma_d v_{td}^2}{n_d}\frac{\partial n_d}{\partial \eta}+\lambda^2\epsilon\nu_{o}v_d=\frac{\eta_o}{n_d}\epsilon\frac{\partial^2v_d}{\partial\xi^2}\\
+\frac{2\lambda\eta_o\epsilon^2}{n_d}\frac{\partial^2v_d}{\partial\xi\partial\eta}
-\frac{\eta_o\epsilon^2}{n_d}\frac{\partial\lambda}{\partial\eta}\frac{\partial n_d}{\partial \xi}+\frac{\eta_o\epsilon^3}{n_d}\frac{\partial^2v_d}{\partial\eta^2},
\end{split}
\end{equation}
\end{widetext}
The Poisson equation is now given as:
\begin{eqnarray}
&\epsilon&\frac{\partial^2\phi}{\partial\xi^2}+2\epsilon^2\lambda\frac{\partial^2\phi}{\partial\xi\partial\eta}
-\epsilon^2\frac{\partial\lambda}{\partial\eta}\frac{\partial\phi}{\partial\xi}
+\epsilon^3\lambda^2\frac{\partial^2\phi}{\partial\eta^2}\\
&+&4\pi\lambda^2\left(n_{io}exp(\frac{-e\phi}{T_i})-n_{eo}exp(\frac{e\phi}{T_e})-Z_dn_d\right)=0.  \nonumber
\end{eqnarray}
We next expand the dependent variables $n_d, v_d$ and $\phi$ in terms of the smallness parameter $\epsilon$ as
\begin{eqnarray}
\psi=\psi_o+\epsilon\psi_1+\epsilon^2\psi_2+...\hspace*{0.1in}.
\end{eqnarray} 
The first order equations in $\epsilon$ of continuity, momentum and Poisson equation leads to the self consistent relation, 
\begin{eqnarray}
\lambda=v_{do}+\sqrt{(C_{da}^2+\gamma_dv_{td}^2)},
\end{eqnarray}
where $\mu=\sqrt{(C_{da}^2+\gamma_dv_{td}^2)}$. In order to derive a modified KdV Burger equation for the in-homogeneous dusty plasma we equate the coefficients of $\epsilon^2$ to zero. The final modified KdV-Burger (m-KdV-Burger) equation governing the shock propagation in a nonuniform dusty plasma  can be expressed by:
 \begin{equation}
 \frac{\partial n_{d1}}{\partial\eta}+A\frac{\partial^3 n_{d1}}{\partial\xi^3}+Bn_{d1}\frac{\partial n_{d1}}{\partial\xi}+C\frac{n_{d1}}{2}=D\frac{\partial^2 n_{d1}}{\partial\xi^2}.
 \label{eq:burger}
 \end{equation}
\begin{figure}[ht]
\includegraphics[scale=0.4]{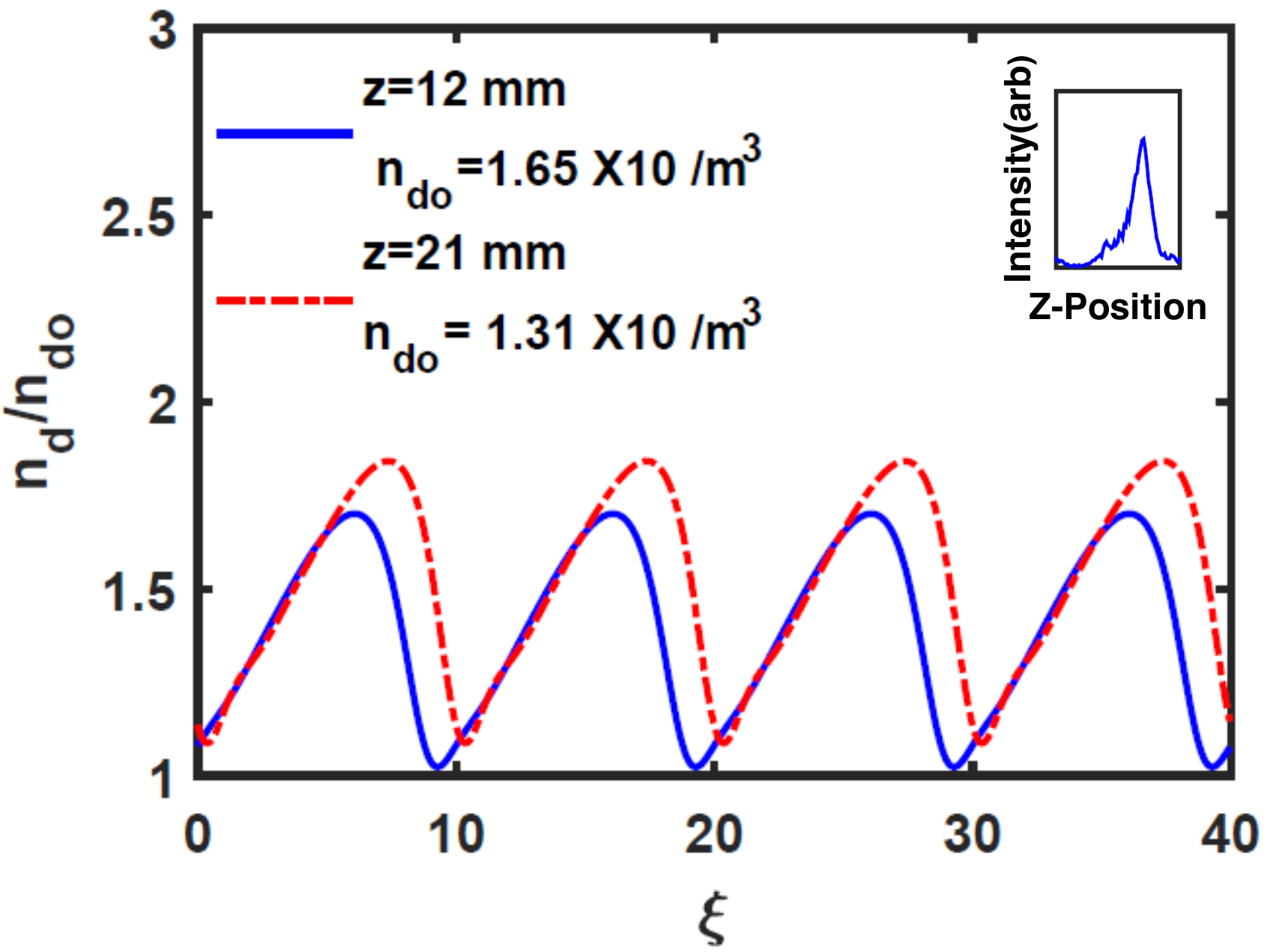}
\caption{\label{fig:fig9} Typical shock wave trains obtained from the numerical solution of m-KdV-Burger equation. Inset shows the same that is obtained experimentally.}
\end{figure}
\textcolor{black}{The terms of Eq.~\ref{eq:burger} having the coefficients $A$, $B$ and $D$ represent dispersive, non-linear and dissipative contributions that normally appear in the KdV Burger equation. The term proportional to \lq C' in the model equation arises due to the dust density inhomogeneity and dust-neutral collisions. The detailed expressions for the coefficients are:
\begin{eqnarray}
&A&=\frac{1}{2n_{do}C_{da}}, \nonumber\\
&B&=\frac{C_{da}^3T_e}{2n_{do}^2Z_de}\left[\frac{1}{Z_d^2}(\sigma_{ei}^2n_{io}-n_{eo})-\frac{3Z_dn_{do}}{C_{da}^4}\right], \nonumber \\
&C&=\frac{1}{2n_{do}}\frac{\partial n_{do}}{\partial\eta}+\frac{\nu_o}{2C_{da}}, \nonumber \\
&D&=\frac{\eta_o}{2C_{da}^3n_{do}}. \nonumber
\end{eqnarray}}
\textcolor{black}{As the background dust density ($n_{do}$) is a function of the axial distance all the coefficients including $C_{da}$ are also functions of the axial distance.} 
{\color{black} In Eq.~\ref{eq:burger} as well as in the above coefficients, normalized quantities are used where  space is normalized by $\sqrt{\frac{\epsilon_0k_BTe}{n_{io}e^2}}$, time is normalized by $\sqrt{\frac{m_d\epsilon_0}{n_{io}e^2}}$, velocity is normalized by $\sqrt{\frac{k_BTe}{m_d}}$ and density is normalized by $n_{io}$. The normalizations are the same as used in references \cite{singh1998linear,singh1999effect} with $\epsilon_0$ and $k_B$ being the permitivity of free space and Boltzmann constant, respectively. While calculating the above coefficients, we have used the experimental values of $T_e=$~4eV, $n_{io}=1.5\times10^{15}$~m$^{-3}$, $Z_d=10^4$, $\nu_0=11$~Hz, $\eta_0=0.2$ \cite{surbhishock} and $\sigma_{ei}=T_e/T_i=133$ for $T_i=0.03$~eV.}  {\color{black}  A general solution  of this equation, even numerically, is quite challenging. So to make some progress and make contact with the experimental results, we have solved this equation locally for constant coefficients at different points of the density profiles and corresponding different values of the coefficients. Fig.~\ref{fig:fig9} shows typical such numerical solutions of Eq.~\ref{eq:burger}, by assuming a uniform dusty plasma with a density value corresponding to the experimental value at $z=12~ mm$ (solid line) and $z=21~ mm$ (dashed line). In other words, the solutions are obtained by calculating the corresponding values of A, B, C and D for these two points and assuming them to remain constant. These plots show that the modified KdV-Burger equation admits  solutions that qualitatively describe the profile of our experimentally observed sawteeth structures and represent nonlinear propagating wave trains \cite{wmtham1974linear} resulting from the balance of nonlinear steepening, nonlinear growth due to the inhomogeneity, dispersive broadening and dissipative damping due to dust-dust correlations and dust neutral collisions.} The inset in Figure~\ref{fig:fig9} shows the experimentally obtained profile of a single shock front extracted from the experimental image shown in Fig.~\ref{fig:fig3}(a). One can see that the numerically obtained profile of a single shock front is very similar to that of the experimental profile.
\textcolor{black}{To see the change in amplitude as well as the width of the sawtooth profile as a function of $z$ we have used such local solutions over the entire range of the density profile, namely from $2.0\times10^{10}$ to $0.7\times10^{10}$~/m$^3$, and plotted the values of the amplitudes and widths in Fig.~\ref{fig:fig10}. The corresponding values of the coefficients  vary in the range $A=0.0383$ to $0.1466$, $B=0.4595$ to $1.7588$, $C=0.4931$ to $0.7648$ and $D=0.0008$ to $0.0073$.}
\begin{figure}[h]
\includegraphics[scale=0.85]{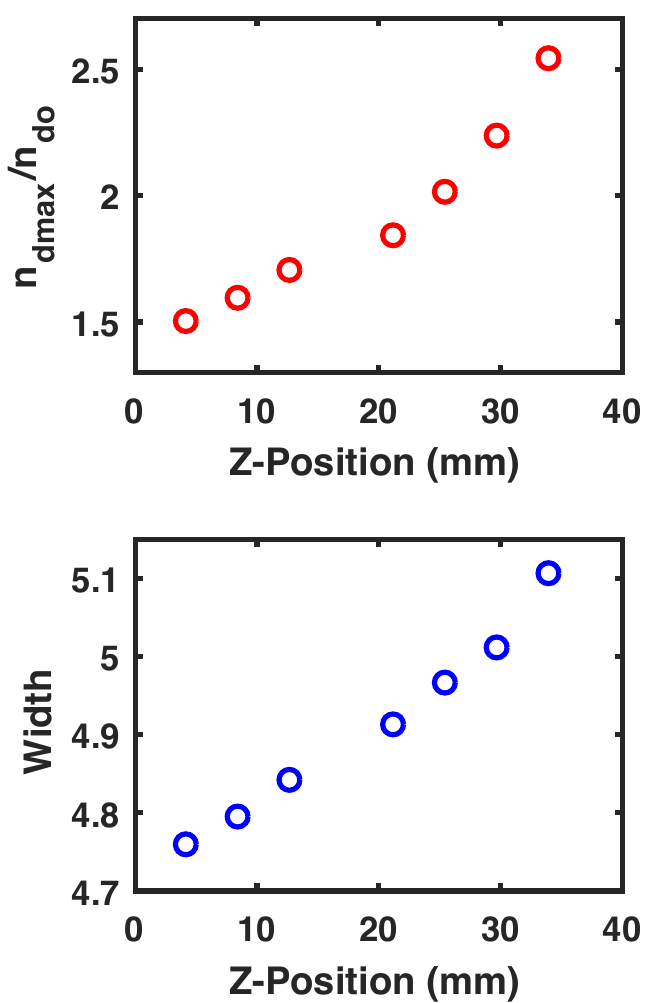}
\caption{\label{fig:fig10} (a) Amplitude and (b) Width profile estimated from the numerical solution of m-KdV Burger equation. }
\end{figure}
One can now clearly see, that as in the experiments, 
the amplitude and width of a single shock structure increases when it propagates down the density gradient.  Thus the m-KdV Burger model provides a good qualitative description of the present experimental results.\par
\section{Summary and Conclusion}\label{sec: summary}
To summarize, we have carried out for the first time a systematic experimental study on the excitation and propagation of non-linear dust acoustic shock waves in an inhomogeneous dusty plasma. The experiments were done in the Dusty Plasma Experimental (DPEx) device in which a direct current glow discharge Argon plasma was formed.  A dusty plasma of kaolin particles was then formed in between an axial confining strip and a grounded copper wire which was installed radially at the midway of the cathode. The inhomogeneity in the dust density was then created by inducing an imbalance of the dynamical equilibrium between the pumping speed and the gas flow rate. A sudden increase of dust density was created by compressing the dust cloud using a single gas injection technique that led to the excitation of a series of nonlinear structures. These structures are characterized by a high compression factor, sharp peaks and a single saw-tooth like profile which are all signatures of a shock wave. The amplitude and the widths of these shock waves are measured along the axial direction from the wire where the equilibrium dust density falls monotonically.  It is found that the amplitude increases and width broadens up as the shock structure propagates down the dust density gradient.  To provide some theoretical understanding of our experimental findings we have developed a model equation in the form of a modified KdV-Burger. The model takes account of the density in inhomogeneity of the dust component, the bulk viscosity due to dust-dust coupling and damping due to dust-neutral collisions.  Numerical solutions of this equation at various density values show spatial profiles as well as variations in the amplitude and width of shock like structures that closely resemble the experimental results. Our results, obtained under controlled laboratory conditions, can help in extending our basic understanding of shock structures in inhomogeneous media and may also find useful applications in the interpretation in related phenomena in astrophysical situations.\\\\
\textbf{Data Availability Statement:} The data that supports the findings of this study are available within the article.   
\begin{acknowledgments}
A.S. is thankful to the Indian National Science Academy (INSA) for their support under the INSA Senior Scientist Fellowship scheme.
\end{acknowledgments}
\vspace*{0.1in}
\noindent\textbf{\large{References}}
\bibliography{ref}

\begin{thebibliography}{38}%
\makeatletter
\providecommand \@ifxundefined [1]{%
 \@ifx{#1\undefined}
}%
\providecommand \@ifnum [1]{%
 \ifnum #1\expandafter \@firstoftwo
 \else \expandafter \@secondoftwo
 \fi
}%
\providecommand \@ifx [1]{%
 \ifx #1\expandafter \@firstoftwo
 \else \expandafter \@secondoftwo
 \fi
}%
\providecommand \natexlab [1]{#1}%
\providecommand \enquote  [1]{``#1''}%
\providecommand \bibnamefont  [1]{#1}%
\providecommand \bibfnamefont [1]{#1}%
\providecommand \citenamefont [1]{#1}%
\providecommand \href@noop [0]{\@secondoftwo}%
\providecommand \href [0]{\begingroup \@sanitize@url \@href}%
\providecommand \@href[1]{\@@startlink{#1}\@@href}%
\providecommand \@@href[1]{\endgroup#1\@@endlink}%
\providecommand \@sanitize@url [0]{\catcode `\\12\catcode `\$12\catcode
  `\&12\catcode `\#12\catcode `\^12\catcode `\_12\catcode `\%12\relax}%
\providecommand \@@startlink[1]{}%
\providecommand \@@endlink[0]{}%
\providecommand \url  [0]{\begingroup\@sanitize@url \@url }%
\providecommand \@url [1]{\endgroup\@href {#1}{\urlprefix }}%
\providecommand \urlprefix  [0]{URL }%
\providecommand \Eprint [0]{\href }%
\providecommand \doibase [0]{http://dx.doi.org/}%
\providecommand \selectlanguage [0]{\@gobble}%
\providecommand \bibinfo  [0]{\@secondoftwo}%
\providecommand \bibfield  [0]{\@secondoftwo}%
\providecommand \translation [1]{[#1]}%
\providecommand \BibitemOpen [0]{}%
\providecommand \bibitemStop [0]{}%
\providecommand \bibitemNoStop [0]{.\EOS\space}%
\providecommand \EOS [0]{\spacefactor3000\relax}%
\providecommand \BibitemShut  [1]{\csname bibitem#1\endcsname}%
\let\auto@bib@innerbib\@empty
\bibitem [{\citenamefont {Barkan}, \citenamefont {Merlino},\ and\ \citenamefont
  {D’Angelo}(1995)}]{dustacousticwave}%
  \BibitemOpen
  \bibfield  {author} {\bibinfo {author} {\bibfnamefont {A.}~\bibnamefont
  {Barkan}}, \bibinfo {author} {\bibfnamefont {R.~L.}\ \bibnamefont {Merlino}},
  \ and\ \bibinfo {author} {\bibfnamefont {N.}~\bibnamefont {D’Angelo}},\
  }\href@noop {} {\bibfield  {journal} {\bibinfo  {journal} {Physics of
  Plasmas}\ }\textbf {\bibinfo {volume} {2}},\ \bibinfo {pages} {3563--3565}
  (\bibinfo {year} {1995})}\BibitemShut {NoStop}%
\bibitem [{\citenamefont {Shukla}\ and\ \citenamefont
  {Silin}(1992)}]{pkshuklaDustionacousticWave}%
  \BibitemOpen
  \bibfield  {author} {\bibinfo {author} {\bibfnamefont {P.~K.}\ \bibnamefont
  {Shukla}}\ and\ \bibinfo {author} {\bibfnamefont {V.~P.}\ \bibnamefont
  {Silin}},\ }\href@noop {} {\bibfield  {journal} {\bibinfo  {journal} {Physica
  Scripta}\ }\textbf {\bibinfo {volume} {45}},\ \bibinfo {pages} {508}
  (\bibinfo {year} {1992})}\BibitemShut {NoStop}%
\bibitem [{\citenamefont {Melandso/}(1996)}]{latticewaves}%
  \BibitemOpen
  \bibfield  {author} {\bibinfo {author} {\bibfnamefont {F.}~\bibnamefont
  {Melandso/}},\ }\href@noop {} {\bibfield  {journal} {\bibinfo  {journal}
  {Physics of Plasmas}\ }\textbf {\bibinfo {volume} {3}},\ \bibinfo {pages}
  {3890--3901} (\bibinfo {year} {1996})}\BibitemShut {NoStop}%
\bibitem [{\citenamefont {Bandyopadhyay}\ \emph
  {et~al.}(2008{\natexlab{a}})\citenamefont {Bandyopadhyay}, \citenamefont
  {Prasad}, \citenamefont {Sen},\ and\ \citenamefont {Kaw}}]{soilton}%
  \BibitemOpen
  \bibfield  {author} {\bibinfo {author} {\bibfnamefont {P.}~\bibnamefont
  {Bandyopadhyay}}, \bibinfo {author} {\bibfnamefont {G.}~\bibnamefont
  {Prasad}}, \bibinfo {author} {\bibfnamefont {A.}~\bibnamefont {Sen}}, \ and\
  \bibinfo {author} {\bibfnamefont {P.~K.}\ \bibnamefont {Kaw}},\ }\href@noop
  {} {\bibfield  {journal} {\bibinfo  {journal} {Phys. Rev. Lett.}\ }\textbf
  {\bibinfo {volume} {101}},\ \bibinfo {pages} {065006} (\bibinfo {year}
  {2008}{\natexlab{a}})}\BibitemShut {NoStop}%
\bibitem [{\citenamefont {Samsonov}\ \emph {et~al.}(2002)\citenamefont
  {Samsonov}, \citenamefont {Ivlev}, \citenamefont {Quinn}, \citenamefont
  {Morfill},\ and\ \citenamefont {Zhdanov}}]{samsonovsoliton}%
  \BibitemOpen
  \bibfield  {author} {\bibinfo {author} {\bibfnamefont {D.}~\bibnamefont
  {Samsonov}}, \bibinfo {author} {\bibfnamefont {A.~V.}\ \bibnamefont {Ivlev}},
  \bibinfo {author} {\bibfnamefont {R.~A.}\ \bibnamefont {Quinn}}, \bibinfo
  {author} {\bibfnamefont {G.}~\bibnamefont {Morfill}}, \ and\ \bibinfo
  {author} {\bibfnamefont {S.}~\bibnamefont {Zhdanov}},\ }\bibfield  {title}
  {\enquote {\bibinfo {title} {Dissipative longitudinal solitons in a
  two-dimensional strongly coupled complex (dusty) plasma},}\ }\href@noop {}
  {\bibfield  {journal} {\bibinfo  {journal} {Phys. Rev. Lett.}\ }\textbf
  {\bibinfo {volume} {88}},\ \bibinfo {pages} {095004} (\bibinfo {year}
  {2002})}\BibitemShut {NoStop}%
\bibitem [{\citenamefont {Sen}\ \emph {et~al.}(2015)\citenamefont {Sen},
  \citenamefont {Tiwari}, \citenamefont {Mishra},\ and\ \citenamefont
  {Kaw}}]{senprecursor}%
  \BibitemOpen
  \bibfield  {author} {\bibinfo {author} {\bibfnamefont {A.}~\bibnamefont
  {Sen}}, \bibinfo {author} {\bibfnamefont {S.}~\bibnamefont {Tiwari}},
  \bibinfo {author} {\bibfnamefont {S.}~\bibnamefont {Mishra}}, \ and\ \bibinfo
  {author} {\bibfnamefont {P.}~\bibnamefont {Kaw}},\ }\href@noop {} {\bibfield
  {journal} {\bibinfo  {journal} {Advances in Space Research}\ }\textbf
  {\bibinfo {volume} {56}},\ \bibinfo {pages} {429 -- 435} (\bibinfo {year}
  {2015})},\ \bibinfo {note} {advances in Asteroid and Space Debris Science and
  Technology - Part 1}\BibitemShut {NoStop}%
\bibitem [{\citenamefont {Heinrich}, \citenamefont {Kim},\ and\ \citenamefont
  {Merlino}(2009)}]{heinrichshock}%
  \BibitemOpen
  \bibfield  {author} {\bibinfo {author} {\bibfnamefont {J.}~\bibnamefont
  {Heinrich}}, \bibinfo {author} {\bibfnamefont {S.-H.}\ \bibnamefont {Kim}}, \
  and\ \bibinfo {author} {\bibfnamefont {R.~L.}\ \bibnamefont {Merlino}},\
  }\bibfield  {title} {\enquote {\bibinfo {title} {Laboratory observations of
  self-excited dust acoustic shocks},}\ }\href@noop {} {\bibfield  {journal}
  {\bibinfo  {journal} {Phys. Rev. Lett.}\ }\textbf {\bibinfo {volume} {103}},\
  \bibinfo {pages} {115002} (\bibinfo {year} {2009})}\BibitemShut {NoStop}%
\bibitem [{\citenamefont {Jaiswal}, \citenamefont {Bandyopadhyay},\ and\
  \citenamefont {Sen}(2016{\natexlab{a}})}]{surbhishock}%
  \BibitemOpen
  \bibfield  {author} {\bibinfo {author} {\bibfnamefont {S.}~\bibnamefont
  {Jaiswal}}, \bibinfo {author} {\bibfnamefont {P.}~\bibnamefont
  {Bandyopadhyay}}, \ and\ \bibinfo {author} {\bibfnamefont {A.}~\bibnamefont
  {Sen}},\ }\href@noop {} {\bibfield  {journal} {\bibinfo  {journal} {Physics
  of Plasmas}\ }\textbf {\bibinfo {volume} {23}},\ \bibinfo {pages} {083701}
  (\bibinfo {year} {2016}{\natexlab{a}})}\BibitemShut {NoStop}%
\bibitem [{\citenamefont {Saitou}\ \emph {et~al.}(2012)\citenamefont {Saitou},
  \citenamefont {Nakamura}, \citenamefont {Kamimura},\ and\ \citenamefont
  {Ishihara}}]{nakamurabow}%
  \BibitemOpen
  \bibfield  {author} {\bibinfo {author} {\bibfnamefont {Y.}~\bibnamefont
  {Saitou}}, \bibinfo {author} {\bibfnamefont {Y.}~\bibnamefont {Nakamura}},
  \bibinfo {author} {\bibfnamefont {T.}~\bibnamefont {Kamimura}}, \ and\
  \bibinfo {author} {\bibfnamefont {O.}~\bibnamefont {Ishihara}},\ }\bibfield
  {title} {\enquote {\bibinfo {title} {Bow shock formation in a complex
  plasma},}\ }\href@noop {} {\bibfield  {journal} {\bibinfo  {journal} {Phys.
  Rev. Lett.}\ }\textbf {\bibinfo {volume} {108}},\ \bibinfo {pages} {065004}
  (\bibinfo {year} {2012})}\BibitemShut {NoStop}%
\bibitem [{\citenamefont {Samsonov}\ and\ \citenamefont
  {Goree}(1999)}]{voidGoree}%
  \BibitemOpen
  \bibfield  {author} {\bibinfo {author} {\bibfnamefont {D.}~\bibnamefont
  {Samsonov}}\ and\ \bibinfo {author} {\bibfnamefont {J.}~\bibnamefont
  {Goree}},\ }\href@noop {} {\bibfield  {journal} {\bibinfo  {journal} {Phys.
  Rev. E}\ }\textbf {\bibinfo {volume} {59}},\ \bibinfo {pages} {1047--1058}
  (\bibinfo {year} {1999})}\BibitemShut {NoStop}%
\bibitem [{\citenamefont {Law}\ \emph {et~al.}(1998)\citenamefont {Law},
  \citenamefont {Steel}, \citenamefont {Annaratone},\ and\ \citenamefont
  {Allen}}]{vortexExcitationLaw}%
  \BibitemOpen
  \bibfield  {author} {\bibinfo {author} {\bibfnamefont {D.~A.}\ \bibnamefont
  {Law}}, \bibinfo {author} {\bibfnamefont {W.~H.}\ \bibnamefont {Steel}},
  \bibinfo {author} {\bibfnamefont {B.~M.}\ \bibnamefont {Annaratone}}, \ and\
  \bibinfo {author} {\bibfnamefont {J.~E.}\ \bibnamefont {Allen}},\ }\href@noop
  {} {\bibfield  {journal} {\bibinfo  {journal} {Phys. Rev. Lett.}\ }\textbf
  {\bibinfo {volume} {80}},\ \bibinfo {pages} {4189--4192} (\bibinfo {year}
  {1998})}\BibitemShut {NoStop}%
\bibitem [{\citenamefont {Usachev}\ \emph {et~al.}(2014)\citenamefont
  {Usachev}, \citenamefont {Zobnin}, \citenamefont {Petrov}, \citenamefont
  {Fortov}, \citenamefont {Thoma}, \citenamefont {Höfner}, \citenamefont
  {Fink}, \citenamefont {Ivlev},\ and\ \citenamefont {Morfill}}]{Usachevshock}%
  \BibitemOpen
  \bibfield  {author} {\bibinfo {author} {\bibfnamefont {A.}~\bibnamefont
  {Usachev}}, \bibinfo {author} {\bibfnamefont {A.}~\bibnamefont {Zobnin}},
  \bibinfo {author} {\bibfnamefont {O.}~\bibnamefont {Petrov}}, \bibinfo
  {author} {\bibfnamefont {V.}~\bibnamefont {Fortov}}, \bibinfo {author}
  {\bibfnamefont {M.~H.}\ \bibnamefont {Thoma}}, \bibinfo {author}
  {\bibfnamefont {H.}~\bibnamefont {Höfner}}, \bibinfo {author} {\bibfnamefont
  {M.}~\bibnamefont {Fink}}, \bibinfo {author} {\bibfnamefont {A.}~\bibnamefont
  {Ivlev}}, \ and\ \bibinfo {author} {\bibfnamefont {G.}~\bibnamefont
  {Morfill}},\ }\bibfield  {title} {\enquote {\bibinfo {title} {Externally
  excited planar dust acoustic shock waves in a strongly coupled dusty plasma
  under microgravity conditions},}\ }\href@noop {} {\bibfield  {journal}
  {\bibinfo  {journal} {New Journal of Physics}\ }\textbf {\bibinfo {volume}
  {16}},\ \bibinfo {pages} {053028} (\bibinfo {year} {2014})}\BibitemShut
  {NoStop}%
\bibitem [{\citenamefont {Steinolfson}\ and\ \citenamefont
  {Cable}(1993)}]{steinolfson1993venus}%
  \BibitemOpen
  \bibfield  {author} {\bibinfo {author} {\bibfnamefont {R.}~\bibnamefont
  {Steinolfson}}\ and\ \bibinfo {author} {\bibfnamefont {S.}~\bibnamefont
  {Cable}},\ }\bibfield  {title} {\enquote {\bibinfo {title} {Venus bow shocks
  at unusually large distances from the planet},}\ }\href@noop {} {\bibfield
  {journal} {\bibinfo  {journal} {Geophysical research letters}\ }\textbf
  {\bibinfo {volume} {20}},\ \bibinfo {pages} {755--758} (\bibinfo {year}
  {1993})}\BibitemShut {NoStop}%
\bibitem [{\citenamefont {Gosling}\ \emph {et~al.}(1967)\citenamefont
  {Gosling}, \citenamefont {Asbridge}, \citenamefont {Bame},\ and\
  \citenamefont {Strong}}]{gosling1967vela}%
  \BibitemOpen
  \bibfield  {author} {\bibinfo {author} {\bibfnamefont {J.}~\bibnamefont
  {Gosling}}, \bibinfo {author} {\bibfnamefont {J.}~\bibnamefont {Asbridge}},
  \bibinfo {author} {\bibfnamefont {S.}~\bibnamefont {Bame}}, \ and\ \bibinfo
  {author} {\bibfnamefont {I.}~\bibnamefont {Strong}},\ }\bibfield  {title}
  {\enquote {\bibinfo {title} {Vela 2 measurements of the magnetopause and bow
  shock positions},}\ }\href@noop {} {\bibfield  {journal} {\bibinfo  {journal}
  {Journal of Geophysical Research}\ }\textbf {\bibinfo {volume} {72}},\
  \bibinfo {pages} {101--112} (\bibinfo {year} {1967})}\BibitemShut {NoStop}%
\bibitem [{\citenamefont {Vidotto}\ \emph {et~al.}(2011)\citenamefont
  {Vidotto}, \citenamefont {Llama}, \citenamefont {Jardine}, \citenamefont
  {Helling},\ and\ \citenamefont {Wood}}]{vidotto2011shock}%
  \BibitemOpen
  \bibfield  {author} {\bibinfo {author} {\bibfnamefont {A.}~\bibnamefont
  {Vidotto}}, \bibinfo {author} {\bibfnamefont {J.}~\bibnamefont {Llama}},
  \bibinfo {author} {\bibfnamefont {M.}~\bibnamefont {Jardine}}, \bibinfo
  {author} {\bibfnamefont {C.}~\bibnamefont {Helling}}, \ and\ \bibinfo
  {author} {\bibfnamefont {K.}~\bibnamefont {Wood}},\ }\bibfield  {title}
  {\enquote {\bibinfo {title} {Shock formation around planets orbiting m-dwarf
  stars},}\ }\href@noop {} {\bibfield  {journal} {\bibinfo  {journal}
  {Astronomische Nachrichten}\ }\textbf {\bibinfo {volume} {332}},\ \bibinfo
  {pages} {1055--1061} (\bibinfo {year} {2011})}\BibitemShut {NoStop}%
\bibitem [{\citenamefont {{Shukla}}\ and\ \citenamefont
  {{Mamun}}(2001)}]{theoretical_shock}%
  \BibitemOpen
  \bibfield  {author} {\bibinfo {author} {\bibfnamefont {P.~K.}\ \bibnamefont
  {{Shukla}}}\ and\ \bibinfo {author} {\bibfnamefont {A.~A.}\ \bibnamefont
  {{Mamun}}},\ }\bibfield  {title} {\enquote {\bibinfo {title} {Dust-acoustic
  shocks in a strongly coupled dusty plasma},}\ }\href@noop {} {\bibfield
  {journal} {\bibinfo  {journal} {IEEE Transactions on Plasma Science}\
  }\textbf {\bibinfo {volume} {29}},\ \bibinfo {pages} {221--225} (\bibinfo
  {year} {2001})}\BibitemShut {NoStop}%
\bibitem [{\citenamefont {Samsonov}\ \emph {et~al.}(2003)\citenamefont
  {Samsonov}, \citenamefont {Morfill}, \citenamefont {Thomas}, \citenamefont
  {Hagl}, \citenamefont {Rothermel}, \citenamefont {Fortov}, \citenamefont
  {Lipaev}, \citenamefont {Molotkov}, \citenamefont {Nefedov}, \citenamefont
  {Petrov}, \citenamefont {Ivanov},\ and\ \citenamefont
  {Krikalev}}]{samsonovshock}%
  \BibitemOpen
  \bibfield  {author} {\bibinfo {author} {\bibfnamefont {D.}~\bibnamefont
  {Samsonov}}, \bibinfo {author} {\bibfnamefont {G.}~\bibnamefont {Morfill}},
  \bibinfo {author} {\bibfnamefont {H.}~\bibnamefont {Thomas}}, \bibinfo
  {author} {\bibfnamefont {T.}~\bibnamefont {Hagl}}, \bibinfo {author}
  {\bibfnamefont {H.}~\bibnamefont {Rothermel}}, \bibinfo {author}
  {\bibfnamefont {V.}~\bibnamefont {Fortov}}, \bibinfo {author} {\bibfnamefont
  {A.}~\bibnamefont {Lipaev}}, \bibinfo {author} {\bibfnamefont
  {V.}~\bibnamefont {Molotkov}}, \bibinfo {author} {\bibfnamefont
  {A.}~\bibnamefont {Nefedov}}, \bibinfo {author} {\bibfnamefont
  {O.}~\bibnamefont {Petrov}}, \bibinfo {author} {\bibfnamefont
  {A.}~\bibnamefont {Ivanov}}, \ and\ \bibinfo {author} {\bibfnamefont
  {S.}~\bibnamefont {Krikalev}},\ }\bibfield  {title} {\enquote {\bibinfo
  {title} {Kinetic measurements of shock wave propagation in a
  three-dimensional complex (dusty) plasma},}\ }\href@noop {} {\bibfield
  {journal} {\bibinfo  {journal} {Phys. Rev. E}\ }\textbf {\bibinfo {volume}
  {67}},\ \bibinfo {pages} {036404} (\bibinfo {year} {2003})}\BibitemShut
  {NoStop}%
\bibitem [{\citenamefont {Xiao}, \citenamefont {Ma},\ and\ \citenamefont
  {Li}(2005)}]{xiao2005dust}%
  \BibitemOpen
  \bibfield  {author} {\bibinfo {author} {\bibfnamefont {D.-l.}\ \bibnamefont
  {Xiao}}, \bibinfo {author} {\bibfnamefont {J.}~\bibnamefont {Ma}}, \ and\
  \bibinfo {author} {\bibfnamefont {Y.-f.}\ \bibnamefont {Li}},\ }\bibfield
  {title} {\enquote {\bibinfo {title} {Dust-acoustic shock waves: Effect of
  plasma density gradient},}\ }\href@noop {} {\bibfield  {journal} {\bibinfo
  {journal} {Physics of plasmas}\ }\textbf {\bibinfo {volume} {12}},\ \bibinfo
  {pages} {052314} (\bibinfo {year} {2005})}\BibitemShut {NoStop}%
\bibitem [{\citenamefont {Zhang}, \citenamefont {Li},\ and\ \citenamefont
  {Du}(2019)}]{zhang2019propagation}%
  \BibitemOpen
  \bibfield  {author} {\bibinfo {author} {\bibfnamefont {L.-P.}\ \bibnamefont
  {Zhang}}, \bibinfo {author} {\bibfnamefont {D.-A.}\ \bibnamefont {Li}}, \
  and\ \bibinfo {author} {\bibfnamefont {H.-M.}\ \bibnamefont {Du}},\
  }\bibfield  {title} {\enquote {\bibinfo {title} {Propagation of shock
  structures in inhomogeneous dusty plasmas with dust size distribution and
  nonadiabatic dust charge fluctuation},}\ }\href@noop {} {\bibfield  {journal}
  {\bibinfo  {journal} {Indian Journal of Physics}\ ,\ \bibinfo {pages} {1--8}}
  (\bibinfo {year} {2019})}\BibitemShut {NoStop}%
\bibitem [{\citenamefont {Tadsen}, \citenamefont {Greiner},\ and\ \citenamefont
  {Piel}(2017)}]{tadsen2017amplitude}%
  \BibitemOpen
  \bibfield  {author} {\bibinfo {author} {\bibfnamefont {B.}~\bibnamefont
  {Tadsen}}, \bibinfo {author} {\bibfnamefont {F.}~\bibnamefont {Greiner}}, \
  and\ \bibinfo {author} {\bibfnamefont {A.}~\bibnamefont {Piel}},\ }\bibfield
  {title} {\enquote {\bibinfo {title} {On the amplitude of dust-density waves
  in inhomogeneous dusty plasmas},}\ }\href@noop {} {\bibfield  {journal}
  {\bibinfo  {journal} {Physics of Plasmas}\ }\textbf {\bibinfo {volume}
  {24}},\ \bibinfo {pages} {033704} (\bibinfo {year} {2017})}\BibitemShut
  {NoStop}%
\bibitem [{\citenamefont {Jaiswal}, \citenamefont {Bandyopadhyay},\ and\
  \citenamefont {Sen}(2015)}]{surbhirsi}%
  \BibitemOpen
  \bibfield  {author} {\bibinfo {author} {\bibfnamefont {S.}~\bibnamefont
  {Jaiswal}}, \bibinfo {author} {\bibfnamefont {P.}~\bibnamefont
  {Bandyopadhyay}}, \ and\ \bibinfo {author} {\bibfnamefont {A.}~\bibnamefont
  {Sen}},\ }\href@noop {} {\bibfield  {journal} {\bibinfo  {journal} {Review of
  Scientific Instruments}\ }\textbf {\bibinfo {volume} {86}},\ \bibinfo {pages}
  {113503} (\bibinfo {year} {2015})}\BibitemShut {NoStop}%
\bibitem [{\citenamefont {Khrapak}\ and\ \citenamefont
  {Morfill}(2006)}]{khrapakcec2}%
  \BibitemOpen
  \bibfield  {author} {\bibinfo {author} {\bibfnamefont {S.~A.}\ \bibnamefont
  {Khrapak}}\ and\ \bibinfo {author} {\bibfnamefont {G.~E.}\ \bibnamefont
  {Morfill}},\ }\bibfield  {title} {\enquote {\bibinfo {title} {Grain surface
  temperature in noble gas discharges: Refined analytical model},}\ }\href@noop
  {} {\bibfield  {journal} {\bibinfo  {journal} {Physics of Plasmas}\ }\textbf
  {\bibinfo {volume} {13}},\ \bibinfo {pages} {104506} (\bibinfo {year}
  {2006})}\BibitemShut {NoStop}%
\bibitem [{\citenamefont {Khrapak}\ \emph {et~al.}(2005)\citenamefont
  {Khrapak}, \citenamefont {Ratynskaia}, \citenamefont {Zobnin}, \citenamefont
  {Usachev}, \citenamefont {Yaroshenko}, \citenamefont {Thoma}, \citenamefont
  {Kretschmer}, \citenamefont {H\"ofner}, \citenamefont {Morfill},
  \citenamefont {Petrov},\ and\ \citenamefont {Fortov}}]{khrapakcec}%
  \BibitemOpen
  \bibfield  {author} {\bibinfo {author} {\bibfnamefont {S.~A.}\ \bibnamefont
  {Khrapak}}, \bibinfo {author} {\bibfnamefont {S.~V.}\ \bibnamefont
  {Ratynskaia}}, \bibinfo {author} {\bibfnamefont {A.~V.}\ \bibnamefont
  {Zobnin}}, \bibinfo {author} {\bibfnamefont {A.~D.}\ \bibnamefont {Usachev}},
  \bibinfo {author} {\bibfnamefont {V.~V.}\ \bibnamefont {Yaroshenko}},
  \bibinfo {author} {\bibfnamefont {M.~H.}\ \bibnamefont {Thoma}}, \bibinfo
  {author} {\bibfnamefont {M.}~\bibnamefont {Kretschmer}}, \bibinfo {author}
  {\bibfnamefont {H.}~\bibnamefont {H\"ofner}}, \bibinfo {author}
  {\bibfnamefont {G.~E.}\ \bibnamefont {Morfill}}, \bibinfo {author}
  {\bibfnamefont {O.~F.}\ \bibnamefont {Petrov}}, \ and\ \bibinfo {author}
  {\bibfnamefont {V.~E.}\ \bibnamefont {Fortov}},\ }\bibfield  {title}
  {\enquote {\bibinfo {title} {Particle charge in the bulk of gas
  discharges},}\ }\href {\doibase 10.1103/PhysRevE.72.016406} {\bibfield
  {journal} {\bibinfo  {journal} {Phys. Rev. E}\ }\textbf {\bibinfo {volume}
  {72}},\ \bibinfo {pages} {016406} (\bibinfo {year} {2005})}\BibitemShut
  {NoStop}%
\bibitem [{\citenamefont {Jaiswal}, \citenamefont {Bandyopadhyay},\ and\
  \citenamefont {Sen}(2016{\natexlab{b}})}]{surbhiflow}%
  \BibitemOpen
  \bibfield  {author} {\bibinfo {author} {\bibfnamefont {S.}~\bibnamefont
  {Jaiswal}}, \bibinfo {author} {\bibfnamefont {P.}~\bibnamefont
  {Bandyopadhyay}}, \ and\ \bibinfo {author} {\bibfnamefont {A.}~\bibnamefont
  {Sen}},\ }\href@noop {} {\bibfield  {journal} {\bibinfo  {journal} {Plasma
  Sources Science and Technology}\ }\textbf {\bibinfo {volume} {25}},\ \bibinfo
  {pages} {065021} (\bibinfo {year} {2016}{\natexlab{b}})}\BibitemShut
  {NoStop}%
\bibitem [{\citenamefont {Arora}\ \emph {et~al.}(2019)\citenamefont {Arora},
  \citenamefont {Bandyopadhyay}, \citenamefont {Hariprasad},\ and\
  \citenamefont {Sen}}]{garimadynamics}%
  \BibitemOpen
  \bibfield  {author} {\bibinfo {author} {\bibfnamefont {G.}~\bibnamefont
  {Arora}}, \bibinfo {author} {\bibfnamefont {P.}~\bibnamefont
  {Bandyopadhyay}}, \bibinfo {author} {\bibfnamefont {M.~G.}\ \bibnamefont
  {Hariprasad}}, \ and\ \bibinfo {author} {\bibfnamefont {A.}~\bibnamefont
  {Sen}},\ }\bibfield  {title} {\enquote {\bibinfo {title} {Micro-dynamics of
  neutral flow induced dusty plasma flow},}\ }\href@noop {} {\bibfield
  {journal} {\bibinfo  {journal} {Physics of Plasmas}\ }\textbf {\bibinfo
  {volume} {26}},\ \bibinfo {pages} {023701} (\bibinfo {year}
  {2019})}\BibitemShut {NoStop}%
\bibitem [{\citenamefont {Merlino}\ \emph {et~al.}(2012)\citenamefont
  {Merlino}, \citenamefont {Heinrich}, \citenamefont {Hyun},\ and\
  \citenamefont {Meyer}}]{merlino2012nonlinear}%
  \BibitemOpen
  \bibfield  {author} {\bibinfo {author} {\bibfnamefont {R.}~\bibnamefont
  {Merlino}}, \bibinfo {author} {\bibfnamefont {J.}~\bibnamefont {Heinrich}},
  \bibinfo {author} {\bibfnamefont {S.-H.}\ \bibnamefont {Hyun}}, \ and\
  \bibinfo {author} {\bibfnamefont {J.}~\bibnamefont {Meyer}},\ }\bibfield
  {title} {\enquote {\bibinfo {title} {Nonlinear dust acoustic waves and
  shocks},}\ }\href@noop {} {\bibfield  {journal} {\bibinfo  {journal} {Physics
  of Plasmas}\ }\textbf {\bibinfo {volume} {19}},\ \bibinfo {pages} {057301}
  (\bibinfo {year} {2012})}\BibitemShut {NoStop}%
\bibitem [{\citenamefont {Whitham}(1974)}]{wmtham1974linear}%
  \BibitemOpen
  \bibfield  {author} {\bibinfo {author} {\bibfnamefont {G.~B.}\ \bibnamefont
  {Whitham}},\ }\href@noop {} {\emph {\bibinfo {title} {Linear and Nonlinear
  Waves}}}\ (\bibinfo  {publisher} {Wiley, New York},\ \bibinfo {year}
  {1974})\BibitemShut {NoStop}%
\bibitem [{\citenamefont {Epstein}(1924)}]{epstein}%
  \BibitemOpen
  \bibfield  {author} {\bibinfo {author} {\bibfnamefont {P.~S.}\ \bibnamefont
  {Epstein}},\ }\href@noop {} {\bibfield  {journal} {\bibinfo  {journal} {Phys.
  Rev.}\ }\textbf {\bibinfo {volume} {23}},\ \bibinfo {pages} {710--733}
  (\bibinfo {year} {1924})}\BibitemShut {NoStop}%
\bibitem [{\citenamefont {Choudhary}, \citenamefont {Mukherjee},\ and\
  \citenamefont {Bandyopadhyay}(2016)}]{choudhary2016propagation}%
  \BibitemOpen
  \bibfield  {author} {\bibinfo {author} {\bibfnamefont {M.}~\bibnamefont
  {Choudhary}}, \bibinfo {author} {\bibfnamefont {S.}~\bibnamefont
  {Mukherjee}}, \ and\ \bibinfo {author} {\bibfnamefont {P.}~\bibnamefont
  {Bandyopadhyay}},\ }\bibfield  {title} {\enquote {\bibinfo {title}
  {Propagation characteristics of dust--acoustic waves in presence of a
  floating cylindrical object in the dc discharge plasma},}\ }\href@noop {}
  {\bibfield  {journal} {\bibinfo  {journal} {Physics of Plasmas}\ }\textbf
  {\bibinfo {volume} {23}},\ \bibinfo {pages} {083705} (\bibinfo {year}
  {2016})}\BibitemShut {NoStop}%
\bibitem [{\citenamefont {Williams}(2012)}]{williams2012spatial}%
  \BibitemOpen
  \bibfield  {author} {\bibinfo {author} {\bibfnamefont {J.~D.}\ \bibnamefont
  {Williams}},\ }\bibfield  {title} {\enquote {\bibinfo {title} {Spatial
  evolution of the dust-acoustic wave},}\ }\href@noop {} {\bibfield  {journal}
  {\bibinfo  {journal} {IEEE Transactions on Plasma Science}\ }\textbf
  {\bibinfo {volume} {41}},\ \bibinfo {pages} {788--793} (\bibinfo {year}
  {2012})}\BibitemShut {NoStop}%
\bibitem [{\citenamefont {Hariprasad}\ \emph {et~al.}(2018)\citenamefont
  {Hariprasad}, \citenamefont {Bandyopadhyay}, \citenamefont {Arora},\ and\
  \citenamefont {Sen}}]{hari2018}%
  \BibitemOpen
  \bibfield  {author} {\bibinfo {author} {\bibfnamefont {M.}~\bibnamefont
  {Hariprasad}}, \bibinfo {author} {\bibfnamefont {P.}~\bibnamefont
  {Bandyopadhyay}}, \bibinfo {author} {\bibfnamefont {G.}~\bibnamefont
  {Arora}}, \ and\ \bibinfo {author} {\bibfnamefont {A.}~\bibnamefont {Sen}},\
  }\bibfield  {title} {\enquote {\bibinfo {title} {Experimental observation of
  a dusty plasma crystal in the cathode sheath of a dc glow discharge
  plasma},}\ }\href@noop {} {\bibfield  {journal} {\bibinfo  {journal} {Physics
  of Plasmas}\ }\textbf {\bibinfo {volume} {25}},\ \bibinfo {pages} {123704}
  (\bibinfo {year} {2018})}\BibitemShut {NoStop}%
\bibitem [{\citenamefont {Annibaldi}\ \emph {et~al.}(2007)\citenamefont
  {Annibaldi}, \citenamefont {Ivlev}, \citenamefont {Konopka}, \citenamefont
  {Ratynskaia}, \citenamefont {Thomas}, \citenamefont {Morfill}, \citenamefont
  {Lipaev}, \citenamefont {Molotkov}, \citenamefont {Petrov},\ and\
  \citenamefont {Fortov}}]{annibaldi2007dust}%
  \BibitemOpen
  \bibfield  {author} {\bibinfo {author} {\bibfnamefont {S.}~\bibnamefont
  {Annibaldi}}, \bibinfo {author} {\bibfnamefont {A.}~\bibnamefont {Ivlev}},
  \bibinfo {author} {\bibfnamefont {U.}~\bibnamefont {Konopka}}, \bibinfo
  {author} {\bibfnamefont {S.}~\bibnamefont {Ratynskaia}}, \bibinfo {author}
  {\bibfnamefont {H.}~\bibnamefont {Thomas}}, \bibinfo {author} {\bibfnamefont
  {G.}~\bibnamefont {Morfill}}, \bibinfo {author} {\bibfnamefont
  {A.}~\bibnamefont {Lipaev}}, \bibinfo {author} {\bibfnamefont
  {V.}~\bibnamefont {Molotkov}}, \bibinfo {author} {\bibfnamefont
  {O.}~\bibnamefont {Petrov}}, \ and\ \bibinfo {author} {\bibfnamefont
  {V.}~\bibnamefont {Fortov}},\ }\bibfield  {title} {\enquote {\bibinfo {title}
  {Dust-acoustic dispersion relation in three-dimensional complex plasmas under
  microgravity},}\ }\href@noop {} {\bibfield  {journal} {\bibinfo  {journal}
  {New Journal of Physics}\ }\textbf {\bibinfo {volume} {9}},\ \bibinfo {pages}
  {327} (\bibinfo {year} {2007})}\BibitemShut {NoStop}%
\bibitem [{\citenamefont {Bandyopadhyay}\ \emph
  {et~al.}(2008{\natexlab{b}})\citenamefont {Bandyopadhyay}, \citenamefont
  {Prasad}, \citenamefont {Sen},\ and\ \citenamefont {Kaw}}]{pintudasw}%
  \BibitemOpen
  \bibfield  {author} {\bibinfo {author} {\bibfnamefont {P.}~\bibnamefont
  {Bandyopadhyay}}, \bibinfo {author} {\bibfnamefont {G.}~\bibnamefont
  {Prasad}}, \bibinfo {author} {\bibfnamefont {A.}~\bibnamefont {Sen}}, \ and\
  \bibinfo {author} {\bibfnamefont {P.~K.}\ \bibnamefont {Kaw}},\ }\bibfield
  {title} {\enquote {\bibinfo {title} {Experimental study of nonlinear dust
  acoustic solitary waves in a dusty plasma},}\ }\href {\doibase
  10.1103/PhysRevLett.101.065006} {\bibfield  {journal} {\bibinfo  {journal}
  {Phys. Rev. Lett.}\ }\textbf {\bibinfo {volume} {101}},\ \bibinfo {pages}
  {065006} (\bibinfo {year} {2008}{\natexlab{b}})}\BibitemShut {NoStop}%
\bibitem [{\citenamefont {Sharma}, \citenamefont {Boruah},\ and\ \citenamefont
  {Bailung}(2014)}]{bailung}%
  \BibitemOpen
  \bibfield  {author} {\bibinfo {author} {\bibfnamefont {S.~K.}\ \bibnamefont
  {Sharma}}, \bibinfo {author} {\bibfnamefont {A.}~\bibnamefont {Boruah}}, \
  and\ \bibinfo {author} {\bibfnamefont {H.}~\bibnamefont {Bailung}},\ }\href
  {\doibase 10.1103/PhysRevE.89.013110} {\bibfield  {journal} {\bibinfo
  {journal} {Phys. Rev. E}\ }\textbf {\bibinfo {volume} {89}},\ \bibinfo
  {pages} {013110} (\bibinfo {year} {2014})}\BibitemShut {NoStop}%
\bibitem [{\citenamefont {Rao}\ and\ \citenamefont
  {Shukla}(1994)}]{rao1994nonlinear}%
  \BibitemOpen
  \bibfield  {author} {\bibinfo {author} {\bibfnamefont {N.}~\bibnamefont
  {Rao}}\ and\ \bibinfo {author} {\bibfnamefont {P.}~\bibnamefont {Shukla}},\
  }\bibfield  {title} {\enquote {\bibinfo {title} {Nonlinear dust-acoustic
  waves with dust charge fluctuations},}\ }\href@noop {} {\bibfield  {journal}
  {\bibinfo  {journal} {Planetary and Space Science}\ }\textbf {\bibinfo
  {volume} {42}},\ \bibinfo {pages} {221--225} (\bibinfo {year}
  {1994})}\BibitemShut {NoStop}%
\bibitem [{\citenamefont {Singh}\ and\ \citenamefont
  {Rao}(1998)}]{singh1998linear}%
  \BibitemOpen
  \bibfield  {author} {\bibinfo {author} {\bibfnamefont {S.}~\bibnamefont
  {Singh}}\ and\ \bibinfo {author} {\bibfnamefont {N.}~\bibnamefont {Rao}},\
  }\bibfield  {title} {\enquote {\bibinfo {title} {Linear and nonlinear
  dust-acoustic waves in inhomogeneous dusty plasmas},}\ }\href@noop {}
  {\bibfield  {journal} {\bibinfo  {journal} {Physics of Plasmas}\ }\textbf
  {\bibinfo {volume} {5}},\ \bibinfo {pages} {94--99} (\bibinfo {year}
  {1998})}\BibitemShut {NoStop}%
\bibitem [{\citenamefont {Singh}\ and\ \citenamefont
  {Rao}(1999)}]{singh1999effect}%
  \BibitemOpen
  \bibfield  {author} {\bibinfo {author} {\bibfnamefont {S.}~\bibnamefont
  {Singh}}\ and\ \bibinfo {author} {\bibfnamefont {N.}~\bibnamefont {Rao}},\
  }\bibfield  {title} {\enquote {\bibinfo {title} {Effect of dust charge
  inhomogeneity on linear and nonlinear dust--acoustic wave propagation},}\
  }\href@noop {} {\bibfield  {journal} {\bibinfo  {journal} {Physics of
  Plasmas}\ }\textbf {\bibinfo {volume} {6}},\ \bibinfo {pages} {3157--3162}
  (\bibinfo {year} {1999})}\BibitemShut {NoStop}%
\bibitem [{\citenamefont {Kaw}\ and\ \citenamefont {Sen}(1998)}]{kaw1998low}%
  \BibitemOpen
  \bibfield  {author} {\bibinfo {author} {\bibfnamefont {P.}~\bibnamefont
  {Kaw}}\ and\ \bibinfo {author} {\bibfnamefont {A.}~\bibnamefont {Sen}},\
  }\bibfield  {title} {\enquote {\bibinfo {title} {Low frequency modes in
  strongly coupled dusty plasmas},}\ }\href@noop {} {\bibfield  {journal}
  {\bibinfo  {journal} {Physics of Plasmas}\ }\textbf {\bibinfo {volume} {5}},\
  \bibinfo {pages} {3552--3559} (\bibinfo {year} {1998})}\BibitemShut {NoStop}%
\end{thebibliography}%
\end{document}